\documentclass[11pt]{article}
\usepackage{amsmath,amsfonts,amssymb,graphicx,epsfig,pdflscape}
\usepackage[usenames]{color}
\usepackage{pstricks}
\numberwithin{equation}{section}

\newcommand{\nc}{\newcommand}
\nc\disp{\displaystyle}
\nc{\fh}{\hat{f}}
\nc{\muh}{\hat{\mu}}
\nc{\nuh}{\hat{\nu}}
\nc{\spos}[2]{\makebox(0,0)[#1]{$\sm{#2}$}}
\nc{\sm}[1]{{\scriptstyle #1}}
\nc{\bib}{\bibitem}
\nc{\al}{\alpha}
\nc{\g}{\gamma}
\nc{\G}{\Gamma}
\nc{\D}{\Delta}
\nc{\eps}{\epsilon}
\nc{\la}{\lambda}
\nc{\La}{\Lambda}
\nc{\var}{\varphi}
\nc{\pa}{\partial}
\nc{\nn}{\nonumber \\ }
\nc{\hf}{\frac{1}{2}}
\nc{\dz}{\frac{dz}{2\pi i}}
\nc{\bin}[2]{\left(\!\!\!\begin{array}{c} {#1}\\ {#2} \end{array}\!\!\!\right)}
\nc{\be}{\begin{equation}}
\nc{\ee}{\end{equation}}
\nc{\bea}{\begin{eqnarray}}
\nc{\eea}{\end{eqnarray}}
\nc{\bra}[1]{\langle {#1}|}
\nc{\ket}[1]{|{#1}\rangle}
\nc{\ketw}[1]{({#1})^{\phantom{a}}_{{\cal W}}}
\nc{\chit}{\raisebox{0.25ex}{$\chi$}}
\nc{\chih}{\raisebox{0.25ex}{$\hat\chi$}}
\nc{\Db}{\mbox{\boldmath $D$}}
\nc{\Hb}{\mbox{\boldmath $H$}}
\nc{\calH}{{\cal H}}
\nc{\calR}{{\cal R}}
\nc{\calL}{{\cal L}}
\nc{\calV}{{\cal V}}
\nc{\Hc}{{\cal H}}
\nc{\Rc}{{\cal R}}
\nc{\Lc}{{\cal L}}
\nc{\Vc}{{\cal V}}
\nc{\Ib}{\mbox{\boldmath $I$}}
\nc{\qb}{\bar{q}}
\nc{\Ac}{\mathcal{A}}
\nc{\Bc}{\mathcal{B}}
\nc{\Cc}{\mathcal{C}}
\nc{\Dc}{\mathcal{D}}
\nc{\Ec}{\mathcal{E}}
\nc{\Ic}{\mathcal{I}}
\nc{\Oc}{\mathcal{O}}
\nc{\Xc}{\mathcal{X}}
\nc{\Yc}{\mathcal{Y}}
\nc{\Zc}{\mathcal{Z}}
\nc{\fus}{\mbox{}\,\hat\otimes\,\mbox{}}
\def\vvdots{\mathinner{\mkern1mu\raise1pt\vbox{\kern7pt\hbox{.}}\mkern2mu
  \raise4pt\hbox{.}\mkern2mu\raise7pt\hbox{.}\mkern1mu}}
\nc{\gauss}[2]{\left[\!\!\begin{array}{c} {#1}\\ {#2} \end{array}\!\!\right]}
\nc{\sbin}[2]{\left\{\!\!\!\begin{array}{c} {#1}\\ {#2} 
\end{array}\!\!\!\right\}}
\nc{\sbinlr}[2]{\Big\langle\!\!\begin{array}{c} {#1}\\ {#2} 
\end{array}\!\!\Big\rangle}
\nc{\bino}[2]{\left(\!\!\begin{array}{c} {#1}\\ {#2} \end{array}\!\!\right)}
\definecolor{lightblue}{rgb}{.61,.61,1}
\definecolor{midblue}{rgb}{.7,.7,1}
\definecolor{lightlightblue}{rgb}{.85,.85,1}
\definecolor{lightestblue}{rgb}{.96,.96,1}
\definecolor{lightpurple}{rgb}{1,.65,1}
\nc{\ch}{{\rm ch}}
\nc{\R}{{\cal R}}
\nc{\dkk}{\delta_{j,\{k,k'\}}^{(2)}}
\nc{\drr}{\delta_{j,\{r,r'\}}^{(2)}}
\nc{\ddkk}{\delta_{j,\{k,k'\}}^{(4)}}
\nc{\dddkk}{\delta_{j,\{k,k'\}}^{(8)}}
\nc{\dnn}{\delta_{j,\{n,n'\}}^{(2)}}
\nc{\ddnn}{\delta_{j,\{n,n'\}}^{(4)}}
\nc{\dddnn}{\delta_{j,\{n,n'\}}^{(8)}}

\definecolor{pink}{rgb}{1,.65,.65}

\begin{document}

\topmargin -5mm
\oddsidemargin 5mm

\setcounter{page}{1}

\vspace{8mm}
\begin{center}
{\LARGE {\bf ${\cal W}$-Extended Fusion Algebra of Critical Percolation}}

\vspace{10mm}
 {\Large J{\o}rgen Rasmussen}\ \ {\large and}\ \ {\Large Paul A. Pearce}
\\[.3cm]
 {\em Department of Mathematics and Statistics, University of Melbourne}\\
 {\em Parkville, Victoria 3010, Australia}
\\[.4cm]
 {\tt J.Rasmussen@ms.unimelb.edu.au}\ \ \ \ \ {\tt P.Pearce@ms.unimelb.edu.au}

\end{center}

\vspace{8mm}
\centerline{{\bf{Abstract}}}
\vskip.4cm
\noindent
Two-dimensional critical percolation is the member ${\cal LM}(2,3)$ of the infinite 
series of Yang-Baxter integrable logarithmic minimal models ${\cal LM}(p,p')$. 
We consider the continuum scaling limit of this lattice model as a `rational' logarithmic conformal 
field theory with extended ${\cal W}={\cal W}_{2,3}$ symmetry and use a lattice approach on a 
strip to study the fundamental fusion rules in this extended picture. We find that the  
representation content of the ensuing closed fusion algebra
contains 26 ${\cal W}$-indecomposable representations with 8 rank-1 representations, 
14 rank-2 representations and 4 rank-3 representations. We identify these representations with 
suitable limits of Yang-Baxter integrable boundary conditions on the lattice and obtain their 
associated ${\cal W}$-extended characters. The latter decompose as finite non-negative sums
of ${\cal W}$-irreducible characters of which 13 are required. 
Implementation of fusion on the lattice allows us 
to read off the fusion rules governing the fusion algebra of the 26 representations and to construct an 
explicit Cayley table. The closure of these representations among themselves under fusion 
is remarkable confirmation of the proposed extended symmetry. 

\renewcommand{\thefootnote}{\arabic{footnote}}
\setcounter{footnote}{0}

\section{Introduction}
\label{SectionIntroduction}

The study of percolation~\cite{BroadHamm57,Kesten82,Grimmet89,Stauffer92} as a lattice model has a long history~\cite{Saleur87,DuplantierSaleur87,SaleurSUSY92}. In this paper, it is convenient to regard two-dimensional critical percolation as the member ${\cal LM}(2,3)$ of the 
infinite series of Yang-Baxter integrable logarithmic minimal models ${\cal LM}(p,p')$~\cite{PRZ}. 
It is a well-established principle that two-dimensional lattice systems in general~\cite{Cardy87} and percolation in particular~\cite{LPS94,Cardy01} are conformally invariant in the continuum scaling limit. Our lattice approach to studying these conformal field theories is predicated on the supposition that, in the continuum scaling limit, a transfer matrix with prescribed boundary conditions gives rise to a representation of the Virasoro algebra. Different boundary conditions naturally lead to different representations which can be of different types --- reducible or irreducible, decomposable or indecomposable. We further assume that, if in addition, the boundary conditions respect the symmetry of a larger conformal algebra ${\cal W}$, then the continuum scaling limit of the transfer matrix will yield a representation of the extended algebra ${\cal W}$.

Notwithstanding the fact that critical percolation is one of the very few systems which has been rigorously shown~\cite{Smirnov01} to be conformally invariant in the continuum scaling limit, the study of critical percolation as a Conformal Field Theory (CFT) is not so well advanced. 
In large part, this is because critical percolation~\cite{Cardy92,Gurarie93,Cardy99,GuLu99,FFHST02,GuLu04,MathieuRidout07}, like critical dense polymers ${\cal LM}(1,2)$~\cite{Gennes, Cloizeaux, Saleur87b, Duplantier, PR07} or symplectic fermions~\cite{Kausch95,Kausch00},  is a prototypical {\em logarithmic}  CFT. The properties~\cite{Flohr03,Gaberdiel03,Kawai03} of logarithmic CFTs differ dramatically from the familiar properties of {\em rational\/} CFTs. In particular, 
they are non-rational and non-unitary with a countably infinite number of scaling fields. Unlike rational CFTs, whose field or representation content consists entirely of {\em irreducible} Virasoro representations, logarithmic CFTs admit {\em reducible yet indecomposable} representations~\cite{Roh96} of the Virasoro algebra. These representations, some of which are accompanied by non-trivial Jordan-cell structures for the Virasoro dilatation generator $L_0$, play an essential role and are in fact characteristic  \mbox{of logarithmic CFTs.}

Recently, Virasoro fusion rules have been 
proposed~\cite{GabKausch96,EberleF06,RS07,RP0706,RP0707} for all the augmented minimal or logarithmic minimal models ${\cal LM}(p,p')$. Interestingly, it was found that only indecomposable representations of rank 1, 2 or 3 appear corresponding to Jordan cells of dimension 1, 2 or 3 respectively. 
However, a central question of much current interest~\cite{Flohr96,GK9606,FG05,GR06} is whether an extended symmetry algebra ${\cal W}$ exists for these logarithmic theories. Such a symmetry should allow the countably {\em infinite} number of Virasoro representations to be reorganized into a {\em finite} number of extended ${\cal W}$-representations which close under fusion. In the case of the logarithmic
minimal models ${\cal LM}(1,p)$, the existence of such an extended ${\cal W}$-symmetry and the associated fusion rules are by now 
well established~\cite{GK9606,FHST03,FGST05,GR07,GTipunin07,PRR08}. 
By stark contrast, although there are strong indications~\cite{FGST06a,FGST06b} that there exists a 
${\cal W}_{p,p'}$ symmetry algebra for general augmented minimal models, very little is known 
about the ${\cal W}$-extended fusion rules for the ${\cal LM}(p,p')$ models with $p\ge 2$.

In this paper, we use a lattice approach on a strip, generalizing the approach of \cite{PRR08}, to obtain fusion rules of critical percolation ${\cal LM}(2,3)$ in the extended symmetry picture. In~\cite{PRR08}, it was shown that in fact symplectic fermions is just critical dense polymers ${\cal LM}(1,2)$ viewed in the extended picture. Likewise in the case of critical percolation, the extended picture is described
by the {\em same} lattice model as the Virasoro picture.
We nevertheless find it useful to distinguish between the two pictures by denoting
the extended picture ${\cal WLM}(2,3)$ and thus reserve the notation ${\cal LM}(2,3)$ for 
critical percolation in the non-extended Virasoro picture. A similar distinction applies to the
entire infinite series of logarithmic minimal models. 
We intend to discuss these ${\cal W}$-extended models, which we denote by ${\cal WLM}(p,p')$, elsewhere.
The ${\cal W}$-extended fusion rules we obtain for critical percolation 
are based on the {\em fundamental\/} fusion algebra in the Virasoro 
picture~\cite{RP0706,RP0707} which is a subset of the {\em full\/} fusion algebra. 
The latter remains to be determined and may eventually yield a larger ${\cal W}$-extended
fusion algebra than the one presented here. 

The layout of this paper is as follows. In Section~2, we review the Virasoro fusion rules for critical percolation \cite{RP0706}. In Section~3, we summarize the ${\cal W}$-representation content consisting of 26 ${\cal W}$-indecomposable representations with 8 rank-1 representations, 14 rank-2 representations and 4 rank-3 representations and present their associated extended characters. 
The latter decompose as finite non-negative sums
of ${\cal W}$-irreducible characters of which 13 are required. These are all identified.
Lastly, in this section, we present the explicit Cayley table of the fundamental ${\cal W}$-extended fusion rules obtained by implementing fusion on the lattice. 
In Section~4, we identify the ${\cal W}$-extended representations with suitable limits of Yang-Baxter integrable boundary conditions on the lattice and give details of their construction and properties. 
We conclude with a short discussion. Throughout, we use the notation $\mathbb{Z}_{n,m}=\mathbb{Z}\cap[n,m]$, with $n,m\in\mathbb{Z}$, to denote the set of integers from $n$ to $m$, both included, and denote an $n$-fold fusion of the representation $A$ with itself by
\be
 A^{\otimes n}=\underbrace{A\otimes A\otimes\ldots\otimes A}_n
\ee

\section{Critical Percolation ${\cal LM}(2,3)$}

\subsection{Logarithmic minimal model ${\cal LM}(p,p')$}

A logarithmic minimal model ${\cal LM}(p,p')$ is defined \cite{PRZ} for every coprime pair of
positive integers $p<p'$.
The model ${\cal LM}(p,p')$ has central charge
\be
 c\ =\  1-6\frac{(p'-p)^2}{pp'}
\label{c}
\ee
and conformal weights
\be
 \D_{r,s}\ =\ \frac{(rp'-sp)^{2}-(p'-p)^2}{4pp'},\hspace{1.2cm}r,s\in\mathbb{N}
\label{D}
\ee
The fundamental fusion algebra $\big\langle(2,1),(1,2)\big\rangle_{p,p'}$ \cite{RP0706,RP0707}
of the logarithmic
minimal model ${\cal LM}(p,p')$ is generated by the two fundamental Kac representations
$(2,1)$ and $(1,2)$ and contains a countably infinite number of inequivalent, indecomposable 
representations of rank 1, 2 or 3. 
For $r,s\in\mathbb{N}$, the character of the Kac representation $(r,s)$ is
\be
 \chit_{r,s}(q)\ =\ \frac{q^{\frac{1-c}{24}+\D_{r,s}}}{\eta(q)}\big(1-q^{rs}\big)
  \ =\ \frac{1}{\eta(q)}\big(q^{(rp'-sp)^2/4pp'}-q^{(rp'+sp)^2/4pp'}\big)
\label{chikac}
\ee
where the Dedekind eta function is given by
\be
  \eta(q)\ =\ q^{\frac{1}{24}} \prod_{n=1}^\infty (1-q^n)
\label{eta}
\ee
Such a representation is of rank 1 and is irreducible if $r\in\mathbb{Z}_{1,p}$ and $s\in p'\mathbb{N}$
or if $r\in p\mathbb{N}$ and $s\in\mathbb{Z}_{1,p'}$. It is a reducible yet indecomposable
representation if $r\in\mathbb{Z}_{1,p-1}$ and $s\in\mathbb{Z}_{1,p'-1}$,
while it is a fully reducible representation if $r\in p\mathbb{N}$ and $s\in p'\mathbb{N}$ where
\be
 (kp,k'p')\ =\ (k'p,kp')\ =\ \bigoplus_{j=|k-k'|+1,\ \!{\rm by}\ \!2}^{k+k'-1}(jp,p')\ =\ 
   \bigoplus_{j=|k-k'|+1,\ \!{\rm by}\ \!2}^{k+k'-1}(p,jp')
\label{kpkp}
\ee
These are the only Kac representations appearing in the fundamental fusion algebra.
The characters of the reducible yet indecomposable Kac representations just mentioned
can be written as sums of two irreducible Virasoro characters
\be
 \chit_{r,s}(q)\ =\ \ch_{r,s}(q)+\ch_{2p-r,s}(q)\ =\ \ch_{r,s}(q)+\ch_{r,2p'-s}(q),\qquad
   r\in\mathbb{Z}_{1,p-1},\quad s\in\mathbb{Z}_{1,p'-1}
\ee
In general and with $r_0\in\mathbb{Z}_{1,p-1}$, $s_0\in\mathbb{Z}_{1,p'-1}$ and $k\in\mathbb{N}-1$, 
the irreducible Virasoro characters read \cite{FSZ}
\bea
 {\rm ch}_{r_0+kp,s_0}(q)\!\!&=&\!\!K_{2pp',(r_0+kp)p'-s_0p;k}(q)-K_{2pp',(r_0+kp)p'+s_0p;k}(q)\nn
 {\rm ch}_{r_0+(k+1)p,p'}(q)\!\!&=&\!\!
  \frac{1}{\eta(q)}\big(q^{(kp+r_0)^2p'/4p}-q^{((k+2)p-r_0)^2p'/4p}\big) \nn
 {\rm ch}_{(k+1)p,s_0}(q)\!\!&=&\!\!
  \frac{1}{\eta(q)}\big(q^{((k+1)p'-s_0)^2p/4p'}-q^{((k+1)p'+s_0)^2p/4p'}\big) \nn
 {\rm ch}_{(k+1)p,p'}(q)\!\!&=&\!\!
  \frac{1}{\eta(q)}\big(q^{k^2pp'/4}-q^{(k+2)^2pp'/4}\big)
\label{laq}
\eea
where $K_{n,\nu;k}(q)$ is defined as
\be
 K_{n,\nu;k}(q)\ =\ \frac{1}{\eta(q)}\sum_{j\in\mathbb{Z}\setminus\mathbb{Z}_{1,k}}q^{(\nu-jn)^2/2n}
\label{Kk}
\ee
For $r\in\mathbb{Z}_{1,p}$, $s\in\mathbb{Z}_{1,p'}$, $a\in\mathbb{Z}_{1,p-1}$, 
$b\in\mathbb{Z}_{1,p'-1}$ and $k\in\mathbb{N}$, 
the representations denoted by $\R_{kp,s}^{a,0}$ and $\R_{r,kp'}^{0,b}$ are indecomposable
representations of rank 2, while $\R_{kp,p'}^{a,b}\equiv\R_{p,kp'}^{a,b}$ 
is an indecomposable representation of rank 3. Their characters read
\bea
 \chit[\R_{kp,s}^{a,0}](q)\!\!&=&\!\!
    \big(1-\delta_{k,1}\delta_{s,p'}\big)\ch_{kp-a,s}(q)+2\ch_{kp+a,s}(q)+\ch_{(k+2)p-a,s}(q)\nn
 \chit[\R_{r,kp'}^{0,b}](q)\!\!&=&\!\!
    \big(1-\delta_{k,1}\delta_{r,p}\big)\ch_{r,kp'-b}(q)+2\ch_{r,kp'+b}(q)+\ch_{r,(k+2)p'-b}(q)\nn
 \chit[\R_{kp,p'}^{a,b}](q)\!\!&=&\!\! \big(1-\delta_{k,1}\big)\ch_{(k-1)p-a,b}(q)+2\ch_{(k-1)p+a,b}(q)
   +2\big(1-\delta_{k,1}\big)\ch_{kp-a,p'-b}(q)\nn
 \!\!&+&\!\!4\ch_{kp+a,p'-b}(q)+\big(2-\delta_{k,1}\big)\ch_{(k+1)p-a,b}(q)
  +2\ch_{(k+1)p+a,b}(q)\nn
 \!\!&+&\!\!2\ch_{(k+2)p-a,p'-b}(q)+\ch_{(k+3)p-a,b}(q)\nn
 \!\!&=&\!\!  \big(1-\delta_{k,1}\big)\ch_{a,(k-1)p'-b}(q)+2\ch_{a,(k-1)p'+b}(q)
   +2\big(1-\delta_{k,1}\big)\ch_{p-a,kp'-b}(q)\nn
 \!\!&+&\!\!4\ch_{p-a,kp'+b}(q)+\big(2-\delta_{k,1}\big)\ch_{a,(k+1)p'-b}(q)
  +2\ch_{a,(k+1)p'+b}(q)\nn
 \!\!&+&\!\!2\ch_{p-a,(k+2)p'-b}(q)+\ch_{a,(k+3)p'-b}(q)
\label{chiR}
\eea
For $a\in\mathbb{Z}_{0,p-1}$, $b\in\mathbb{Z}_{0,p'-1}$ and $k,k'\in\mathbb{N}$, 
a decomposition similar to (\ref{kpkp}) applies to the higher-rank {\em decomposable}
representations $\R_{kp,k'p'}^{a,b}$ as we have 
\be
 \R_{kp,k'p'}^{a,b}\ =\ \R_{k'p,kp'}^{a,b}
   \ =\ \bigoplus_{j=|k-k'|+1,\ \!{\rm by}\ \!2}^{k+k'-1}\R_{jp,p'}^{a,b}
   \ =\ \bigoplus_{j=|k-k'|+1,\ \!{\rm by}\ \!2}^{k+k'-1}\R_{p,jp'}^{a,b}
\ee
Here we have introduced the convenient notation
\be
 \R_{r,s}^{0,0}\ \equiv\ (r,s)
\ee
Fusion in the fundamental fusion algebra $\big\langle(2,1),(1,2)\big\rangle_{p,p'}$
decomposes into `horizontal' and `vertical' components. With 
$a\in\mathbb{Z}_{0,p-1}$, $b\in\mathbb{Z}_{0,p'-1}$ and $k\in\mathbb{N}$, we thus have 
\be
 \R_{p,kp'}^{a,b}\ =\ \R_{p,1}^{a,0}\otimes\R_{1,kp'}^{0,b}\ =\ \R_{kp,1}^{a,0}\otimes\R_{1,p'}^{0,b}
\label{decomp}
\ee
The Kac representation $(1,1)$ is the identity of the fundamental fusion algebra.
For $p>1$, this is a reducible yet indecomposable representation, while for $p=1$, it
is an irreducible representation.
Below, we summarize the fusion rules in the case of critical percolation ${\cal LM}(2,3)$.
The associated extended Kac table is given in Figure \ref{KacTable}.
%
%
\begin{figure}[p]
\psset{unit=1.cm}
{\small
\begin{center}
\qquad\qquad
\begin{pspicture}(0,0)(7,11)
\psframe[linewidth=0pt,fillstyle=solid,fillcolor=lightestblue](0,0)(7,11)
\psframe[linewidth=1pt,fillstyle=solid,fillcolor=lightblue](0,0)(1,2)
\psframe[linewidth=0pt,fillstyle=solid,fillcolor=lightlightblue](1,0)(2,11)
\psframe[linewidth=0pt,fillstyle=solid,fillcolor=lightlightblue](3,0)(4,11)
\psframe[linewidth=0pt,fillstyle=solid,fillcolor=lightlightblue](5,0)(6,11)
\psframe[linewidth=0pt,fillstyle=solid,fillcolor=lightlightblue](0,2)(7,3)
\psframe[linewidth=0pt,fillstyle=solid,fillcolor=lightlightblue](0,5)(7,6)
\psframe[linewidth=0pt,fillstyle=solid,fillcolor=lightlightblue](0,8)(7,9)
\multiput(0,0)(0,3){3}{\multiput(0,0)(2,0){3}{\psframe[linewidth=0pt,fillstyle=solid,fillcolor=midblue](1,2)(2,3)}}
\multirput(2,1)(2,0){3}{\pswedge[fillstyle=solid,fillcolor=red,linecolor=red](0,0){.25}{180}{270}}
\multirput(2,2)(2,0){3}{\pswedge[fillstyle=solid,fillcolor=red,linecolor=red](0,0){.25}{180}{270}}
\multirput(2,3)(2,0){3}{\pswedge[fillstyle=solid,fillcolor=red,linecolor=red](0,0){.25}{180}{270}}
\multirput(1,3)(0,3){3}{\pswedge[fillstyle=solid,fillcolor=red,linecolor=red](0,0){.25}{180}{270}}
\multirput(2,3)(0,3){3}{\pswedge[fillstyle=solid,fillcolor=red,linecolor=red](0,0){.25}{180}{270}}
\psgrid[gridlabels=0pt,subgriddiv=1]
\rput(.5,10.65){$\vdots$}\rput(1.5,10.65){$\vdots$}\rput(2.5,10.65){$\vdots$}\rput(3.5,10.65){$\vdots$}\rput(4.5,10.65){$\vdots$}\rput(5.5,10.65){$\vdots$}\rput(6.5,10.5){$\vvdots$}
\rput(.5,9.5){$12$}\rput(1.5,9.5){$\frac{65}8$}\rput(2.5,9.5){$5$}\rput(3.5,9.5){$\frac{21}8$}\rput(4.5,9.5){$1$}\rput(5.5,9.5){$\frac{1}8$}\rput(6.5,9.5){$\cdots$}
\rput(.5,8.5){$\frac{28}3$}\rput(1.5,8.5){$\frac{143}{24}$}\rput(2.5,8.5){$\frac{10}3$}\rput(3.5,8.5){$\frac{35}{24}$}\rput(4.5,8.5){$\frac 13$}\rput(5.5,8.5){$-\frac{1}{24}$}\rput(6.5,8.5){$\cdots$}
\rput(.5,7.5){$7$}\rput(1.5,7.5){$\frac {33}8$}\rput(2.5,7.5){$2$}\rput(3.5,7.5){$\frac{5}8$}\rput(4.5,7.5){$0$}\rput(5.5,7.5){$\frac{1}8$}\rput(6.5,7.5){$\cdots$}
\rput(.5,6.5){$5$}\rput(1.5,6.5){$\frac {21}8$}\rput(2.5,6.5){$1$}\rput(3.5,6.5){$\frac{1}8$}\rput(4.5,6.5){$0$}\rput(5.5,6.5){$\frac{5}8$}\rput(6.5,6.5){$\cdots$}
\rput(.5,5.5){$\frac{10}3$}\rput(1.5,5.5){$\frac {35}{24}$}\rput(2.5,5.5){$\frac 13$}\rput(3.5,5.5){$-\frac{1}{24}$}\rput(4.5,5.5){$\frac 13$}\rput(5.5,5.5){$\frac{35}{24}$}\rput(6.5,5.5){$\cdots$}
\rput(.5,4.5){$2$}\rput(1.5,4.5){$\frac 58$}\rput(2.5,4.5){$0$}\rput(3.5,4.5){$\frac{1}8$}\rput(4.5,4.5){$1$}\rput(5.5,4.5){$\frac{21}8$}\rput(6.5,4.5){$\cdots$}
\rput(.5,3.5){$1$}\rput(1.5,3.5){$\frac 18$}\rput(2.5,3.5){$0$}\rput(3.5,3.5){$\frac{5}8$}\rput(4.5,3.5){$2$}\rput(5.5,3.5){$\frac{33}8$}\rput(6.5,3.5){$\cdots$}
\rput(.5,2.5){$\frac 13$}\rput(1.5,2.5){$-\frac 1{24}$}\rput(2.5,2.5){$\frac 13$}\rput(3.5,2.5){$\frac{35}{24}$}\rput(4.5,2.5){$\frac{10}3$}\rput(5.5,2.5){$\frac{143}{24}$}\rput(6.5,2.5){$\cdots$}
\rput(.5,1.5){$0$}\rput(1.5,1.5){$\frac 18$}\rput(2.5,1.5){$1$}\rput(3.5,1.5){$\frac{21}8$}\rput(4.5,1.5){$5$}\rput(5.5,1.5){$\frac{65}8$}\rput(6.5,1.5){$\cdots$}
\rput(.5,.5){$0$}\rput(1.5,.5){$\frac 58$}\rput(2.5,.5){$2$}\rput(3.5,.5){$\frac{33}8$}\rput(4.5,.5){$7$}\rput(5.5,.5){$\frac{85}8$}\rput(6.5,.5){$\cdots$}
{\color{blue}
\rput(.5,-.5){$1$}
\rput(1.5,-.5){$2$}
\rput(2.5,-.5){$3$}
\rput(3.5,-.5){$4$}
\rput(4.5,-.5){$5$}
\rput(5.5,-.5){$6$}
\rput(6.5,-.5){$r$}
\rput(-.5,.5){$1$}
\rput(-.5,1.5){$2$}
\rput(-.5,2.5){$3$}
\rput(-.5,3.5){$4$}
\rput(-.5,4.5){$5$}
\rput(-.5,5.5){$6$}
\rput(-.5,6.5){$7$}
\rput(-.5,7.5){$8$}
\rput(-.5,8.5){$9$}
\rput(-.5,9.5){$10$}
\rput(-.5,10.5){$s$}}
\end{pspicture}
\end{center}}
\caption{Extended Kac table of critical percolation ${\cal LM}(2,3)$ showing the conformal weights 
$\Delta_{r,s}$ of the Kac representations $(r,s)$ where $r,s\in\mathbb{N}$. 
Except for the identifications $(2k,3k')=(2k',3k)$, the entries relate to 
{\em distinct} Kac representations even if the conformal weights coincide. This is unlike the
irreducible representations which are uniquely characterized by their conformal weight.  
The Kac representations which happen to be irreducible representations are marked with a 
red-shaded quadrant in the top-right corner. These do not exhaust the distinct values of the conformal weights. For example, the irreducible representation with $\Delta_{1,1}=0$ does not arise as a Kac representation. By contrast, the Kac table of the associated {\em rational} (minimal) model consisting of the shaded $1\times 2$ grid in the lower-left corner is trivial and contains only the operator corresponding to the irreducible representation with $\D=0$.}
\label{KacTable}
\end{figure}

\subsection{Fundamental fusion algebra of ${\cal LM}(2,3)$}

The fundamental fusion algebra $\big\langle(2,1),(1,2)\big\rangle_{2,3}$ is generated by
the irreducible Kac representation $(2,1)$ and the reducible yet indecomposable Kac representation
$(1,2)$ and contains a variety of representations
\be
  \big\langle(2,1), (1,2)\big\rangle_{2,3}\ =\ \big\langle(1,1), (1,2), (2k,s), (r,3k),
   \R_{2k,s}^{1,0}, \R_{r,3k}^{0,b},  
  \R_{2k,3}^{1,b}\big\rangle_{2,3}
\label{A2112}
\ee
where $r,b\in\mathbb{Z}_{1,2}$, $s\in\mathbb{Z}_{1,3}$ and $k\in\mathbb{N}$.
The representations $(2k,3)\equiv(2,3k)$ are listed twice and it is recalled that 
$\R_{2k,3}^{1,0}\equiv\R_{2,3k}^{1,0}$, $\R_{2k,3}^{0,b}\equiv\R_{2,3k}^{0,b}$
and $\R_{2k,3}^{1,b}\equiv\R_{2,3k}^{1,b}$.
As already mentioned, the reducible yet indecomposable Kac representation $(1,1)$
is the identity of the fundamental fusion algebra
\be
 (1,1)\otimes A\ =\ A
\ee
where $A$ is any of the representations listed in (\ref{A2112}).
Thanks to the decomposition illustrated in (\ref{decomp}), the fundamental fusion
algebra follows from a straightforward merge of the horizontal and vertical components.
To appreciate this, we follow \cite{RP0706} and let 
$A_{r,s}=\bar{a}_{r,1}\otimes\ a_{1,s}$, $B_{r',s'}=\bar{b}_{r',1}\otimes\ b_{1,s'}$,
$\bar{a}_{r,1}\otimes\bar{b}_{r',1}=\bigoplus_{r''}\bar{c}_{r'',1}$ and
$a_{1,s}\otimes b_{1,s'}=\bigoplus_{s''}c_{1,s''}$. Our fusion prescription now yields
\bea
 A_{r,s}\otimes B_{r',s'}\!\!&=&\!\!\Big(\bar{a}_{r,1}\otimes a_{1,s}\Big)\otimes
  \Big(\bar{b}_{r',1}\otimes b_{1,s'}\Big)
   \ =\ \Big(\bar{a}_{r,1}\otimes\bar{b}_{r',1}\Big)\otimes
  \Big(a_{1,s}\otimes b_{1,s'}\Big)\nn
 \!\!&=&\!\!\Big(\bigoplus_{r''}\bar{c}_{r'',1}\Big)\otimes\Big(\bigoplus_{s''}c_{1,s''}\Big)
  \ =\ \bigoplus_{r'',s''}C_{r'',s''}
\label{rs}
\eea
where $C_{r'',s''}=\bar{c}_{r'',1}\otimes c_{1,s''}$. In order to describe the component
fusion algebras explicitly, we introduce the Kronecker delta combinations \cite{RP0706}
\bea
 \dkk\!\!&=&\!\!  2-\delta_{j,|k-k'|}-\delta_{j,k+k'}  \nn
 \ddkk\!\!&=&\!\!   4-3\delta_{j,|k-k'|-1}-2\delta_{j,|k-k'|}-\delta_{j,|k-k'|+1}
   -\delta_{j,k+k'-1}-2\delta_{j,k+k'}-3\delta_{j,k+k'+1}
\label{d24}
\eea
where $k,k'\in\mathbb{N}$.
The horizontal fusion algebra
\be
 \big\langle(2,1)\big\rangle_{2,3}\ =\ \big\langle(2k,1),\R_{2k,1}^{1,0}\big\rangle_{2,3}
\ee 
then reads
\bea
 (2k,1)\otimes(2k',1)\!\!&=&\!\!\bigoplus_{j=|k-k'|+1,\ \!{\rm by}\ \!2}^{k+k'-1}
  \R_{2j,1}^{1,0}\nn
(2k,1)\otimes \R_{2k',1}^{1,0}\!\!&=&\!\!\bigoplus_{j=|k-k'|}^{k+k'}
  \dkk(2j,1)\nn
 \R_{2k,1}^{1,0}\otimes \R_{2k',1}^{1,0}\!\!&=&\!\!\bigoplus_{j=|k-k'|}^{k+k'}
  \dkk\R_{2j,1}^{1,0}
   \label{fusion21}
\eea
while the vertical fusion algebra 
\be
 \big\langle(1,2)\big\rangle_{2,3}
    \ =\ \big\langle(1,1),(1,2),(1,3k),\R_{1,3k}^{0,1},\R_{1,3k}^{0,2}\big\rangle_{2,3}
\label{A12}
\ee
reads
\bea
 (1,1)\otimes A\!\!&=&\!\!A\nn
 (1,2)\otimes (1,2)\!\!&=&\!\!(1,1)\oplus(1,3)\nn
 (1,2)\otimes (1,3k)\!\!&=&\!\!\R_{1,3k}^{0,1}\nn
 (1,2)\otimes \R_{1,3k}^{0,1}\!\!&=&\!\!\R_{1,3k}^{0,2}\oplus2(1,3k)\nn
 (1,2)\otimes \R_{1,3k}^{0,2}\!\!&=&\!\!\R_{1,3k}^{0,1}\oplus(1,3(k-1))\oplus (1,3(k+1))\nn
 (1,3k)\otimes(1,3k')\!\!&=&\!\!
  \bigoplus_{j=|k-k'|+1,\ \!{\rm by}\ \!2}^{k+k'-1}\big(\R_{1,3j}^{0,2}\oplus(1,3j)\big)\nn
 (1,3k)\otimes \R_{1,3k'}^{0,1}\!\!&=&\!\!
   \Big(\bigoplus_{j=|k-k'|+1,\ \!{\rm by}\ \!2}^{k+k'-1}2\R_{1,3j}^{0,1}\Big)
  \oplus\Big(\bigoplus_{j=|k-k'|,\ \!{\rm by}\ \!2}^{k+k'}\dkk (1,3j)\Big)\nn
 (1,3k)\otimes \R_{1,3k'}^{0,2}\!\!&=&\!\!  
  \Big(\bigoplus_{j=|k-k'|,\ \!{\rm by}\ \!2}^{k+k'}\dkk\R_{1,3j}^{0,1}\Big)
  \oplus\Big(\bigoplus_{j=|k-k'|+1,\ \!{\rm by}\ \!2}^{k+k'-1}2(1,3j)\Big)\nn
 \R_{1,3k}^{0,1}\otimes \R_{1,3k'}^{0,1}\!\!&=&\!\!\ 
  \Big(\bigoplus_{j=|k-k'|,\ \!{\rm by}\ \!2}^{k+k'}\dkk\R_{1,3j}^{0,1}\Big)
  \oplus\Big(\bigoplus_{j=|k-k'|+1,\ \!{\rm by}\ \!2}^{k+k'-1}\big(2\R_{1,3j}^{0,2}\oplus4(1,3j)\big)\Big)\nn
 \R_{1,3k}^{0,1}\otimes \R_{1,3k'}^{0,2}\!\!&=&\!\!
  \Big(\bigoplus_{j=|k-k'|+1,\ \!{\rm by}\ \!2}^{k+k'-1}2\R_{1,3j}^{0,1}\Big)
  \oplus 
   \Big(\bigoplus_{j=|k-k'|,\ \!{\rm by}\ \!2}^{k+k'}\dkk\big(\R_{1,3j}^{0,2}
   \oplus2(1,3j)\big)\Big)\nn
 \R_{1,3k}^{0,2}\otimes R_{1,3k'}^{0,2}\!\!&=&\!\! 
 \Big(\bigoplus_{j=|k-k'|,\ \!{\rm by}\ \!2}^{k+k'}\dkk\R_{1,3j}^{0,1}\Big)
  \oplus\Big(\bigoplus_{j=|k-k'|+1,\ \!{\rm by}\ \!2}^{k+k'-1}2\R_{1,3j}^{0,2}\Big)\nn
\!\!&\oplus&\!\! \Big(\bigoplus_{j=|k-k'|-1,\ \!{\rm by}\ \!2}^{k+k'+1}
  \ddkk(1,3j)\Big)
\label{fusion12}
\eea
where $A$ is any of the representations listed in (\ref{A12}).
To illustrate the merge of the two components, we conclude this discussion of critical percolation in the
Virasoro picture ${\cal LM}(2,3)$ by considering the fusion 
\bea
 \R_{2k,3}^{1,1}\otimes \R_{2k',3}^{1,1}\!\!&=&\!\!\Big(\R_{2k,1}^{1,0}\otimes \R_{1,3}^{0,1}\Big)
   \otimes\Big(\R_{2k',1}^{1,0}\otimes \R_{1,3}^{0,1}\Big)
   \ =\ \Big(\R_{2k,1}^{1,0}\otimes \R_{2k',1}^{1,0}\Big)
    \otimes\Big(\R_{1,3}^{0,1}\otimes \R_{1,3}^{0,1}\Big)\nn
 \!\!&=&\!\!\Big(\bigoplus_{j=|k-k'|}^{k+k'}\dkk\R_{2j,1}^{1,0}\Big)
   \otimes\Big(\R_{1,6}^{0,1}\oplus2\R_{1,3}^{0,2}\oplus4(1,3)\Big)\nn
 \!\!&=&\!\!\Big(\bigoplus_{j=|k-k'|-1}^{k+k'+1}\ddkk\R_{2j,3}^{1,1}\Big)
    \oplus\Big(\bigoplus_{j=|k-k'|}^{k+k'}\dkk\big(2\R_{2j,3}^{1,2}\oplus4\R_{2j,3}^{1,0}\big)\Big)
\label{ex3311}
\eea

\section{${\cal W}$-Extended Critical Percolation ${\cal WLM}(2,3)$}

In this section, we summarize our findings in the extended picture for the representation content, 
their characters and their closed fusion algebra. 
Unless otherwise specified, we let $\kappa,r,b\in\mathbb{Z}_{1,2}$,
$s\in\mathbb{Z}_{1,3}$ and $k,k'\in\mathbb{N}$ in the following.

\subsection{Summary of representation content}

We have the 8 ${\cal W}$-indecomposable rank-1 representations
\be
 \big\{\ketw{2\kappa,s},\ketw{r,3\kappa}\big\}
 \hspace{1.2cm}\mathrm{subject\ to}\ \ \ \ketw{2,6}\equiv\ketw{4,3}
\label{8r1}
\ee
where $\ketw{2,3}$ is listed twice,
the 14 ${\cal W}$-indecomposable rank-2 representations
\be
 \big\{\ketw{\R_{2\kappa,s}^{1,0}}, \ketw{\R_{r,3\kappa}^{0,b}}\big\}
\label{14r2}
\ee
and the 4 ${\cal W}$-indecomposable rank-3 representations
\be
 \big\{\ketw{\R_{2,3}^{1,b}},\ketw{\R_{2,6}^{1,b}},\ketw{\R_{4,3}^{1,b}}\big\}
 \hspace{1.2cm}\mathrm{subject\ to}\ \ \ \ketw{\R_{2,6}^{1,b}}\equiv\ketw{\R_{4,3}^{1,b}}
\label{4r3}
\ee
Here we are asserting that these ${\cal W}$-representations are indeed ${\cal W}$-indecomposable.
In terms of Virasoro-indecomposable representations,
the ${\cal W}$-indecomposable rank-1 representations decompose as
\bea
 \ketw{2\kappa,s}&=&\bigoplus_{k\in\mathbb{N}}(2k-2+\kappa)(2(2k-2+\kappa),s)\nn
 \ketw{r,3\kappa}&=&\bigoplus_{k\in\mathbb{N}}(2k-2+\kappa)(r,3(2k-2+\kappa))
\eea
where the two expressions for $\ketw{2,3}$ agree and where
\be
 \ketw{2,6}\ \equiv\ \ketw{4,3}
\ee
Likewise, the ${\cal W}$-indecomposable rank-2 representations decompose as
\bea
 \ketw{\R_{2\kappa,s}^{1,0}}&=&
   \bigoplus_{k\in\mathbb{N}}(2k-2+\kappa)\R_{2(2k-2+\kappa),s}^{1,0}\nn
 \ketw{\R_{r,3\kappa}^{0,b}}&=&\bigoplus_{k\in\mathbb{N}}(2k-2+\kappa)\R_{r,3(2k-2+\kappa)}^{0,b}
\label{Rr2}
\eea
Finally, the ${\cal W}$-indecomposable rank-3 representations decompose as
\bea
 \ketw{\R_{2\kappa,3}^{1,b}}
   &=&\bigoplus_{k\in\mathbb{N}}(2k-2+\kappa)\R_{2(2k-2+\kappa),3}^{1,b}\nn
 \ketw{\R_{2,3\kappa}^{1,b}}
    &=&\bigoplus_{k\in\mathbb{N}}(2k-2+\kappa)\R_{2,3(2k-2+\kappa)}^{1,b}
\label{Rr3}
\eea
where the two expressions for $\ketw{\R_{2,3}^{1,b}}$ agree and where
\be
 \ketw{\R_{2,6}^{1,b}}\ \equiv\ \ketw{\R_{4,3}^{1,b}}
\ee

\subsection{Summary of ${\cal W}$-extended characters}

The characters of the ${\cal W}$-indecomposable rank-1 representations read
\be
 \chih_{2\kappa,s}(q)\ =\ \sum_{k\in\mathbb{N}}(2k-2+\kappa)\ch_{2(2k-2+\kappa),s}(q),\qquad
   \chih_{r,3\kappa}(q)\ =\ \sum_{k\in\mathbb{N}}(2k-2+\kappa)\ch_{r,3(2k-2+\kappa)}(q)
\label{chihch}
\ee
where it is recalled that $\ketw{4,3}\equiv\ketw{2,6}$.
The characters of the ${\cal W}$-indecomposable rank-2 representations read
\bea
 \chit[\ketw{\R_{2\kappa,s}^{1,0}}](q)
   &=&\delta_{\kappa,1}\big\{1-\delta_{s,3}\big\}+\sum_{k\in\mathbb{N}}4k\,\ch_{4k+1,s}(q)
   +\sum_{k\in\mathbb{N}}(4k-2)\ch_{4k-1,s}(q)  \nn
 \chit[\ketw{\R_{r,3\kappa}^{0,b}}](q)
   &=&\delta_{\kappa,1}\big\{1-\delta_{r,2}\big\}
    +\sum_{k\in\mathbb{N}}(4k+2-2\kappa)\ch_{r,6k+6-3\kappa-b}(q)\nn
  &&\hspace{1.2cm}+\sum_{k\in\mathbb{N}}(4k-4+2\kappa)\ch_{r,6k-6+3\kappa+b}(q) 
\eea
We note the character identities
\be
 \chit[\ketw{\R_{2,3}^{1,0}}](q)\ =\ \chit[\ketw{\R_{4,3}^{1,0}}](q),\qquad
 \chit[\ketw{\R_{2,3}^{0,b}}](q)\ =\ \chit[\ketw{\R_{2,6}^{0,3-b}}](q)
\label{chid1}
\ee
and the character relations
\be
  \chit[\ketw{\R_{2,b}^{1,0}}](q)\ =\ 1+\chit[\ketw{\R_{4,b}^{1,0}}](q),\qquad
    \chit[\ketw{\R_{1,3}^{0,b}}](q)\ =\ 1+\chit[\ketw{\R_{1,6}^{0,3-b}}](q)
\label{chrel1}
\ee
and
\be
 \chit[\ketw{\R_{1,3\kappa}^{0,1}}](q)+\chit[\ketw{\R_{1,3\kappa}^{0,2}}](q)
   \ =\ \chit[\ketw{\R_{2\kappa,1}^{1,0}}](q)+\chit[\ketw{\R_{2\kappa,2}^{1,0}}](q)
\label{chrelsum}
\ee
The characters of the ${\cal W}$-indecomposable rank-3 representations read
\be
 \chit[\ketw{\R_{2\kappa,3}^{1,b}}](q)\ =\ 2+\sum_{k\in\mathbb{N}}4k\,\ch_{2k+1,b}(q)
   +\sum_{k\in\mathbb{N}}8k\,\ch_{4k+1,3-b}(q)+\sum_{k\in\mathbb{N}}(8k-4)
     \ch_{4k-1,3-b}(q)
\label{chitRr3}
\ee
and are seen to be independent of $\kappa$. As we will discuss below, 
the dependence on $\kappa$ manifests itself in the distinct Jordan-cell and general embedding
structures of $\ketw{\R_{2\kappa,3}^{1,b}}$ for different $\kappa,b\in\mathbb{Z}_{1,2}$. 
Likewise, the ${\cal W}$-indecomposable
rank-2 representations appearing in (\ref{chid1}) have distinct embedding structures.

We also have ${\cal W}$-extended characters of the various {\em subfactors} of the 
${\cal W}$-indecomposable representations
\bea
 \chih_{0}(q)&=& 1\nn
 \chih_{1}(q)&=& \sum_{k\in\mathbb{N}}(2k-1)\ch_{4k-1,2}(q)\ =\ 
   \sum_{k\in\mathbb{N}}(2k-1)\ch_{1,6k-2}(q)\nn
   &=&\frac{1}{\eta(q)}\sum_{k\in\mathbb{Z}} k^2\Big(q^{(12k-7)^2/24}-q^{(12k+1)^2/24}\Big)\nn
 \chih_{2}(q)&=&   \sum_{k\in\mathbb{N}}(2k-1)\ch_{4k-1,1}(q)\ =\ 
   \sum_{k\in\mathbb{N}}(2k-1)\ch_{1,6k-1}(q)\nn
  &=&\frac{1}{\eta(q)}\sum_{k\in\mathbb{Z}} k^2\Big(q^{(12k-5)^2/24}-q^{(12k-1)^2/24}\Big)\nn
 \chih_{5}(q)&=&   \sum_{k\in\mathbb{N}}2k\,\ch_{4k+1,2}(q)\ =\ 
   \sum_{k\in\mathbb{N}}2k\,\ch_{1,6k+1}(q)\nn
   &=&\frac{1}{\eta(q)}\sum_{k\in\mathbb{Z}} k(k+1)\Big(q^{(12k-1)^2/24}-q^{(12k+7)^2/24}\Big)\nn
 \chih_{7}(q)&=&   \sum_{k\in\mathbb{N}}2k\,\ch_{4k+1,1}(q)\ =\ 
   \sum_{k\in\mathbb{N}}2k\,\ch_{1,6k+2}(q)\nn
  &=&\frac{1}{\eta(q)}\sum_{k\in\mathbb{Z}} k(k+1)\Big(q^{(12k+1)^2/24}-q^{(12k+5)^2/24}\Big)
\label{chih}
\eea
Here we have used the notation $\chih_{\D}(q)$, where $\D$ is the conformal dimension of
the corresponding representation, and some of the identities
\bea
 &&\D_{1,6k+2}\ =\ \D_{4k+1,1},\qquad\D_{1,6k+1}\ =\ \D_{4k+1,2},\qquad\D_{1,6k-1}\ =\ \D_{4k-1,1},
   \qquad\D_{1,6k-2}\ =\ \D_{4k-1,2}\nn
 &&\D_{2,6k+2}\ =\ \D_{4k,1},\quad\qquad\D_{2,6k+1}\ =\ \D_{4k,2},
   \quad\qquad\D_{2,6k-1}\ =\ \D_{4k-2,1},
   \qquad\D_{2,6k-2}\ =\ \D_{4k-2,2}\nn
  &&\D_{1,3k}\ =\ \D_{2k+1,3},\quad\qquad\D_{2,3k}\ =\ \D_{2k,3}
\eea
Similarly, written as $\chih_{\D}(q)$, the 8 independent characters in (\ref{chihch}) read
\bea
 &&\chih_{\frac{1}{3}}(q)\ =\ \sum_{k\in\mathbb{N}}(2k-1)\ch_{4k-1,3}(q)
   \ =\ \sum_{k\in\mathbb{N}}(2k-1)\ch_{1,6k-3}(q)
   \ =\ \frac{1}{\eta(q)} \sum_{k\in\mathbb{Z}} (2k-1)q^{3(4k-3)^2/8}\nn
  &&\chih_{\frac{10}{3}}(q)\ =\ \sum_{k\in\mathbb{N}}2k\,\ch_{4k+1,3}(q)
    \ =\ \sum_{k\in\mathbb{N}}2k\,\ch_{1,6k}(q)
    \ =\ \frac{1}{\eta(q)} \sum_{k\in\mathbb{Z}} 2k\,q^{3(4k-1)^2/8}
\label{chihsum1}
\eea
and
\bea
 &&\chih_{\frac{1}{8}}(q)\ =\ \sum_{k\in\mathbb{N}}(2k-1)\ch_{4k-2,2}(q)
   \ =\ \sum_{k\in\mathbb{N}}(2k-1)\ch_{2,6k-2}(q)
    \ =\ \frac{1}{\eta(q)} \sum_{k\in\mathbb{Z}} (2k-1)\,q^{(6k-5)^2/6}\nn
 &&\chih_{\frac{5}{8}}(q)\ =\ \sum_{k\in\mathbb{N}}(2k-1)\ch_{4k-2,1}(q)
   \ =\ \sum_{k\in\mathbb{N}}(2k-1)\ch_{2,6k-1}(q)
    \ =\ \frac{1}{\eta(q)} \sum_{k\in\mathbb{Z}} (2k-1)\,q^{(6k-4)^2/6}\nn
 &&\chih_{\frac{21}{8}}(q)\ =\ \sum_{k\in\mathbb{N}}2k\,\ch_{4k,2}(q)
   \ =\ \sum_{k\in\mathbb{N}}2k\,\ch_{2,6k+1}(q)
   \ =\ \frac{1}{\eta(q)} \sum_{k\in\mathbb{Z}} 2k\,q^{(6k-2)^2/6}\nn
 &&\chih_{\frac{33}{8}}(q)\ =\ \sum_{k\in\mathbb{N}}2k\,\ch_{4k,1}(q)
   \ =\ \sum_{k\in\mathbb{N}}2k\,\ch_{2,6k+2}(q)
   \ =\ \frac{1}{\eta(q)} \sum_{k\in\mathbb{Z}} 2k\,q^{(6k-1)^2/6}
\label{chihsum2}
\eea
and
\bea
 &&\chih_{-\frac{1}{24}}(q)\ =\ \sum_{k\in\mathbb{N}}(2k-1)\ch_{4k-2,3}(q)
   \ =\ \sum_{k\in\mathbb{N}}(2k-1)\ch_{2,6k-3}(q)
   \ =\ \frac{1}{\eta(q)} \sum_{k\in\mathbb{Z}} (2k-1)\,q^{(6k-6)^2/6}\nn
 &&\chih_{\frac{35}{24}}(q)\ =\ \sum_{k\in\mathbb{N}}2k\,\ch_{4k,3}(q)
   \ =\ \sum_{k\in\mathbb{N}}2k\,\ch_{2,6k}(q) 
    \ =\ \frac{1}{\eta(q)} \sum_{k\in\mathbb{Z}} 2k\,q^{(6k-3)^2/6}
\label{chihsum3}
\eea
We believe that the 5 characters in (\ref{chih}) and the 8 characters in (\ref{chihsum1})
through (\ref{chihsum3})
correspond to ${\cal W}$-{\em irreducible} representations. This yields a total of
13 ${\cal W}$-irreducible representations.
In terms of these irreducible characters, we have the decompositions
\bea
 \chit[\ketw{\R_{2,3}^{1,0}}](q)&=& \chit[\ketw{\R_{4,3}^{1,0}}](q)
   \ =\ 2\chih_{\frac{1}{3}}(q)+2\chih_{\frac{10}{3}}(q)\nn
 \chit[\ketw{\R_{2,3}^{0,1}}](q)&=&\chit[\ketw{\R_{2,6}^{0,2}}](q)
   \ =\ 2\chih_{\frac{1}{8}}(q)+2\chih_{\frac{33}{8}}(q)\nn
 \chit[\ketw{\R_{2,3}^{0,2}}](q)&=&\chit[\ketw{\R_{2,6}^{0,1}}](q)
   \ =\  2\chih_{\frac{5}{8}}(q)+2\chih_{\frac{21}{8}}(q)
\label{chiR2irrrat}
\eea
and
\bea
 \chit[\ketw{\R_{2,1}^{1,0}}](q)&=&1+\chit[\ketw{\R_{4,1}^{1,0}}](q)
   \ =\ 1+2\chih_{2}(q)+2\chih_{7}(q)\nn
 \chit[\ketw{\R_{2,2}^{1,0}}](q)&=&1+\chit[\ketw{\R_{4,2}^{1,0}}](q)
   \ =\ 1+2\chih_{1}(q)+2\chih_{5}(q)\nn
 \chit[\ketw{\R_{1,3}^{0,1}}](q)&=&1+\chit[\ketw{\R_{1,6}^{0,2}}](q)
   \ =\ 1+2\chih_{1}(q)+2\chih_{7}(q)\nn
 \chit[\ketw{\R_{1,3}^{0,2}}](q)&=&1+\chit[\ketw{\R_{1,6}^{0,1}}](q)
   \ =\ 1+2\chih_{2}(q)+2\chih_{5}(q)
\label{chiR2irr}
\eea
and
\be
 \chit[\ketw{\R_{2\kappa,3}^{1,b}}](q)\ =\ 2+4\chih_{1}(q)+4\chih_{2}(q)+4\chih_{5}(q)+4\chih_{7}(q) 
\label{chiR3irr}
\ee
The ${\cal W}$-irreducible representations whose
characters are given by (\ref{chih}) are denoted below by $\ketw{\D}$. 
Sometimes, we extend this practice to the 
${\cal W}$-irreducible representations (\ref{chihch}) as well.
We refer to the finite Kac table in Figure \ref{FiniteKacTable} for a natural organization of
the conformal weights of the 13 ${\cal W}$-irreducible representations.

Letting $\chit_{r,s}(q)$ denote the character of the Kac representation $(r,s)$ where $r,s\in\mathbb{N}$,
we have
\bea
 \chih_0(q)&=&\chit_{1,1}(q)-\sum_{k\in\mathbb{N}}\Big(\chit_{4k-1,1}(q)-\chit_{4k+1,1}(q)\Big)
  \ =\ \chit_{1,1}(q)-\sum_{k\in\mathbb{N}}\Big(\chit_{1,6k-1}(q)-\chit_{1,6k+1}(q)\Big)\nn
 \chih_1(q)&=&\sum_{k\in\mathbb{N}}k^2\Big(\chit_{4k-1,2}(q)-\chit_{4k+1,2}(q)\Big)
  \ =\ \sum_{k\in\mathbb{N}}k^2\Big(\chit_{1,6k-2}(q)-\chit_{1,6k+2}(q)\Big)\nn
 \chih_2(q)&=&\sum_{k\in\mathbb{N}}k^2\Big(\chit_{4k-1,1}(q)-\chit_{4k+1,1}(q)\Big)
  \ =\ \sum_{k\in\mathbb{N}}k^2\Big(\chit_{1,6k-1}(q)-\chit_{1,6k+1}(q)\Big)\\
 \chih_5(q)&=&\sum_{k\in\mathbb{N}}k(k+1)\Big(\chit_{4k+1,2}(q)-\chit_{4(k+1)-1,2}(q)\Big)
  \ =\ \sum_{k\in\mathbb{N}}k(k+1)\Big(\chit_{1,6k+1}(q)-\chit_{1,6(k+1)-1}(q)\Big)\nn
 \chih_7(q)&=&\sum_{k\in\mathbb{N}}k(k+1)\Big(\chit_{4k+1,1}(q)-\chit_{4(k+1)-1,1}(q)\Big)
  \ =\ \sum_{k\in\mathbb{N}}k(k+1)\Big(\chit_{1,6k+2}(q)-\chit_{1,6(k+1)-2}(q)\Big)\nonumber
\label{chihKac}
\eea
Since the Kac representations appearing in (\ref{chihch}) and (\ref{chihsum1}) through
(\ref{chihsum3}) are {\em irreducible} Virasoro representations themselves, we have 
\be
 \chit_{2(2k-2+\kappa),s}(q)\ =\ \ch_{2(2k-2+\kappa),s}(q),\qquad
 \chit_{r,3(2k-2+\kappa)}(q)\ =\ \ch_{r,3(2k-2+\kappa)}(q)
\ee
and hence 
\be
 \chih_{2\kappa,s}(q)\ =\ \sum_{k\in\mathbb{N}}(2k-2+\kappa)\chit_{2(2k-2+\kappa),s}(q),\qquad
   \chih_{r,3\kappa}(q)\ =\ \sum_{k\in\mathbb{N}}(2k-2+\kappa)\chit_{r,3(2k-2+\kappa)}(q)
\ee
\begin{figure}[p]
\psset{unit=1.35cm}
\setlength{\unitlength}{1.35cm}
{
\begin{center}
\begin{pspicture}(0,-.3)(5,8)
\psframe[linewidth=0pt,fillstyle=solid,fillcolor=lightestblue](0,0)(5,8)
\psframe[linewidth=1pt,fillstyle=solid,fillcolor=lightblue](0,0)(1,2)
\psframe[linewidth=0pt,fillstyle=solid,fillcolor=lightlightblue](1,0)(2,8)
\psframe[linewidth=0pt,fillstyle=solid,fillcolor=lightlightblue](3,0)(4,8)
\psframe[linewidth=0pt,fillstyle=solid,fillcolor=lightlightblue](5,0)(5,8)
\psframe[linewidth=0pt,fillstyle=solid,fillcolor=lightlightblue](0,2)(5,3)
\psframe[linewidth=0pt,fillstyle=solid,fillcolor=lightlightblue](0,5)(5,6)
\psframe[linewidth=0pt,fillstyle=solid,fillcolor=lightlightblue](0,8)(5,8)
\multiput(0,0)(0,3){2}{\multiput(0,0)(2,0){2}{\psframe[linewidth=0pt,fillstyle=solid,fillcolor=midblue](1,2)(2,3)}}
\psframe[linewidth=0pt,fillstyle=solid,fillcolor=pink](2,3)(5,8)
\multirput(2,1)(2,0){2}{\pswedge[fillstyle=solid,fillcolor=red,linecolor=red](0,0){.25}{180}{270}}
\multirput(2,2)(2,0){2}{\pswedge[fillstyle=solid,fillcolor=red,linecolor=red](0,0){.25}{180}{270}}
\multirput(2,3)(2,0){2}{\pswedge[fillstyle=solid,fillcolor=red,linecolor=red](0,0){.25}{180}{270}}
\multirput(1,3)(0,3){2}{\pswedge[fillstyle=solid,fillcolor=red,linecolor=red](0,0){.25}{180}{270}}
\multirput(2,3)(0,3){2}{\pswedge[fillstyle=solid,fillcolor=red,linecolor=red](0,0){.25}{180}{270}}
\psgrid[gridlabels=0pt,subgriddiv=1](0,0)(5,8)
\rput(.5,7.5){$7$}\rput(1.5,7.5){$\frac {33}8$}
\rput(.5,6.5){$5$}\rput(1.5,6.5){$\frac {21}8$}
\rput(.5,5.5){$\frac{10}3$}\rput(1.5,5.5){$\frac {35}{24}$}
\rput(.5,4.5){$2$}\rput(1.5,4.5){$\frac 58$}
\rput(.5,3.5){$1$}\rput(1.5,3.5){$\frac 18$}
\rput(.5,2.5){$\frac 13$}\rput(1.5,2.5){$-\frac 1{24}$}\rput(2.5,2.5){$\frac 13$}\rput(3.5,2.5){$\frac{35}{24}$}\rput(4.5,2.5){$\frac{10}3$}
\rput(.5,1.5){$0$}\rput(1.5,1.5){$\frac 18$}\rput(2.5,1.5){$1$}\rput(3.5,1.5){$\frac{21}8$}\rput(4.5,1.5){$5$}
\rput(.5,.5){$0$}\rput(1.5,.5){$\frac 58$}\rput(2.5,.5){$2$}\rput(3.5,.5){$\frac{33}8$}\rput(4.5,.5){$7$}
{\color{blue}
\rput(.5,-.5){$1$}
\rput(1.5,-.5){$2$}
\rput(2.5,-.5){$3$}
\rput(3.5,-.5){$4$}
\rput(4.5,-.5){$5$}
\rput(5.5,-.5){$r$}
\rput(-.5,.5){$1$}
\rput(-.5,1.5){$2$}
\rput(-.5,2.5){$3$}
\rput(-.5,3.5){$4$}
\rput(-.5,4.5){$5$}
\rput(-.5,5.5){$6$}
\rput(-.5,6.5){$7$}
\rput(-.5,7.5){$8$}
\rput(-.5,8.5){$s$}}
\end{pspicture}
\end{center}}
\caption{
Finite part of the infinite Kac table of critical percolation. This part, which is relevant in the extended picture ${\cal WLM}(2,3)$, 
corresponds to the bottom-left corner of the infinite Kac table of Figure~1.
}
\label{FiniteKacTable}
\end{figure}
\begin{figure}
\psset{unit=1.35cm}
\setlength{\unitlength}{1.35cm}{
\small\begin{center}
\begin{pspicture}(-1,-.3)(3,3.5)
 \psframe[linewidth=0pt,fillstyle=solid,fillcolor=lightestblue](0,0)(2,3)
 \psframe[linewidth=1pt,fillstyle=solid,fillcolor=lightblue](0,0)(1,2)
 \psframe[linewidth=0pt,fillstyle=solid,fillcolor=lightlightblue](1,0)(2,3)
 \psframe[linewidth=0pt,fillstyle=solid,fillcolor=lightlightblue](0,2)(2,3)
 \multiput(0,0)(0,3){1}{\multiput(0,0)(2,0){1}{
    \psframe[linewidth=0pt,fillstyle=solid,fillcolor=midblue](1,2)(2,3)}}
 \psgrid[gridlabels=0pt,subgriddiv=1](0,0)(2,3)
 \rput(.5,2.5){$\frac 13,\frac{10}3$}
 \rput(1.5,2.5){$-\frac 1{24},\frac{35}{24}$}
 \rput(.5,1.5){$1,5$}
 \rput(1.5,1.5){$\frac 18,\frac{21}8$}
 \rput(.5,.5){$(0)\,2,7$}
 \rput(1.5,.5){$\frac 58,\frac{33}8$}
{\color{blue}
 \rput(.5,-.5){$1$}
 \rput(1.5,-.5){$2$}
 \rput(2.5,-.5){$r$}
 \rput(-.5,.5){$1$}
 \rput(-.5,1.5){$2$}
 \rput(-.5,2.5){$3$}
 \rput(-.5,3.5){$s$}}
\end{pspicture}
\end{center}}
\caption{Schematic finite Kac table, following \cite{FGST06b}, of the 13 ${\cal W}$-irreducible representations for critical percolation in the extended picture ${\cal WLM}(2,3)$.}\label{WirredKac}
\label{RussianKacTable}
\end{figure}

\subsubsection{Theta forms}

The characters of the 13 ${\cal W}$-irreducible representations agree with those of \cite{FGST06b}.Ê
In particular, they admit the expressions given there in terms of theta functions
\be
  \theta_{\ell,k}(q,z)\ =\ \sum_{j\in\mathbb{Z}+\frac{\ell}{2k}} 
    q^{kj^2} z^{k j},\qquad |q|<1,\quad z\in\mathbb{C},\quad k\in\mathbb{N},\quad \ell\in\mathbb{Z}
\ee
and theta-constants
\be
  \theta_{\ell,k}(q)\ =\ \theta_{\ell,k}(q,1),\quad \theta_{\ell,k}^{(m)}(q)
    \ =\ \bigg(z\frac{\partial}{\partial z}\bigg)^m\theta_{\ell,k}(q,z)\bigg|_{z=1},\qquad m\in\mathbb{N}
\ee
Introducing the abbreviations
\be
  \theta_\ell(q)\ =\ \theta_{\ell,pp'}(q),\qquad \theta'_\ell(q)\ =\ \theta_{\ell,pp'}^{(1)}(q),\qquad 
    \theta''_\ell(q)\ =\ \theta^{(2)}_{\ell,pp'}(q)
\ee
the theta forms are
\bea
\chih_{r,s}(q)&=&\frac{1}{\eta(q)}(\theta_{sp-rp'}(q)-\theta_{sp+rp'}(q)),
    \qquad r\in\mathbb{Z}_{1,p-1},\quad s\in\mathbb{Z}_{1,p'-1}, \quad sp+rp'\le pp'\\
 \chih^+_{r,s}(q)&=&\frac{1}{(pp')^2\eta(q)}\bigg(\theta''_{sp+rp'}(q)
   -\theta''_{sp-rp'}(q)-(sp+rp')\theta'_{sp+rp'}(q)+(sp-rp')\theta'_{sp-rp'}(q)\nonumber\\
 &&\mbox{}+\frac{(sp+rp')^2}{4}\,\theta_{sp+rp'}(q)-\frac{(sp-rp')^2}{4}\,\theta_{sp-rp'}(q)\bigg),
   \qquad r\in\mathbb{Z}_{1,p},\quad s\in\mathbb{Z}_{1,p'}\\
  \chih^-_{r,s}(q)&=&\frac{1}{(pp')^2\eta(q)}\bigg(\theta''_{pp'-sp-rp'}(q)
     -\theta''_{pp'+sp-rp'}(q)+(sp+rp')\theta'_{pp'-sp-rp'}(q)\nonumber\\
  &&\mbox{}+(sp-rp')\theta'_{pp'+sp-rp'}(q)+\frac{(sp+rp')^2-(pp')^2}{4}\,\theta_{pp'-sp-rp'}(q)
   \nonumber\\
 &&\mbox{}-\frac{(sp-rp')^2-(pp')^2}{4}\,\theta_{pp'+sp-rp'}(q)\bigg),
   \qquad r\in\mathbb{Z}_{1,p},\quad s\in\mathbb{Z}_{1,p'}
\eea
where the Dedekind eta function is defined in (\ref{eta}).
As the notation suggests, these are believed to be the theta forms relevant in the case of general $p,p'$
\cite{FGST06b}.
It is noted that the theta form $\chih_{r,s}(q)$ is identical to the well-known irreducible Virasoro
character $\chit_{\D_{r,s}}(q)=\ch_{r,s}(q)$.
The precise relations between our ${\cal W}$-irreducible characters 
and the theta forms for $p=2$ and $p'=3$ are
\bea
 \chih_0(q)\ =\ {\chih}_{1,1}(q)\ =\ 1,\qquad\begin{array}{ll}
    \chih_1(q)\ =\ \chih^+_{1,2}(q),\qquad 
  &\chih_5(q)\ =\ \chih^-_{1,2}(q)\\[10pt]
    \chih_2(q)\ =\ \chih^+_{1,1}(q),\qquad 
  &\chih_7(q)\ =\ \chih^-_{1,1}(q)\end{array}\\[6pt]
  \chih_{2\kappa,s}(q)\ =\ \begin{cases}\chih^+_{2,s}(q),
   &\kappa=1\\ 
    \chih^-_{2,s}(q),
   &\kappa=2\end{cases}\qquad\qquad
   \chih_{r,3\kappa}(q)\ =\ \begin{cases}\chih^+_{r,3}(q),
   &\kappa=1\\ 
    \chih^-_{r,3}(q),
   &\kappa=2\end{cases}\quad\mbox{}
\eea
The ${\cal W}$-irreducible characters $\chih_{2,3}(q)$ and $\chih_{4,3}(q)=\chih_{2,6}(q)$
are listed twice. A compact version of the Kac table in Figure \ref{FiniteKacTable} is given in 
Figure \ref{RussianKacTable}.

\subsection{Embedding diagrams and Jordan-cell structures}

We conjecture that every ${\cal W}$-indecomposable rank-2 representation
has an embedding pattern of one of the types 
\be
 \mbox{
 \begin{picture}(100,120)(0,0)
    \unitlength=1cm
  \thinlines
\put(-2.3,2){$\mathcal{E}(\D_h,\D_v):$}
\put(2,3.6){$\ketw{\D_v}$}
\put(0,2){$\ketw{\D_h}$}
\put(3.9,2){$\ketw{\D_h}$}
\put(2,0.5){$\ketw{\D_v}$}
\put(3.5,2.1){\vector(-1,0){2.1}}
\put(1.9,3.4){\vector(-4,-3){1.2}}
\put(4.1,1.7){\vector(-4,-3){1.2}}
\put(1.9,0.8){\vector(-4,3){1.2}}
\put(4.1,2.5){\vector(-4,3){1.2}}
 \end{picture}
}
\hspace{5.3cm}
 \mbox{
 \begin{picture}(100,120)(-10,0)
    \unitlength=1cm
  \thinlines
\put(-3,2){$\mathcal{E}(\D_h,\D_v;\D_c):$}
\put(2,3.6){$\ketw{\D_v}$}
\put(0,2){$\ketw{\D_h}$}
\put(3.9,2){$\ketw{\D_h}$}
\put(2,0.5){$\ketw{\D_v}$}
\put(3.5,2.1){\vector(-1,0){2.1}}
\put(1.9,3.4){\vector(-4,-3){1.2}}
\put(4.1,1.7){\vector(-4,-3){1.2}}
\put(1.9,0.8){\vector(-4,3){1.2}}
\put(4.1,2.5){\vector(-4,3){1.2}}
\put(2,1.35){$\ketw{\D_c}$}
\put(3.7,1.9){\vector(-2,-1){0.5}}
\put(1.8,1.6){\vector(-2,1){0.5}}
 \end{picture}
}
\label{E}
\ee 
where the horizontal arrows indicate the non-diagonal action of the Virasoro mode $L_0$.
Specifically, we conjecture that the 14 ${\cal W}$-indecomposable rank-2 representations 
(\ref{14r2}) enjoy the embedding patterns 
\bea
 &&\mbox{}\hspace{-1cm}\ketw{\R_{2,1}^{1,0}}\sim\mathcal{E}(2,7;0)
   ,\qquad \ketw{\R_{4,1}^{1,0}}\sim\mathcal{E}(7,2),\qquad\qquad
   \ketw{\R_{2,3}^{1,0}}\sim\mathcal{E}(\frac{1}{3},\frac{10}{3})
   ,\qquad \ketw{\R_{4,3}^{1,0}}\sim\mathcal{E}(\frac{10}{3},\frac{1}{3})\nn
 &&\mbox{}\hspace{-1cm}\ketw{\R_{2,2}^{1,0}}\sim\mathcal{E}(1,5;0)
   ,\qquad \ketw{\R_{4,2}^{1,0}}\sim\mathcal{E}(5,1),\qquad\qquad
   \ketw{\R_{2,3}^{0,1}}\sim\mathcal{E}(\frac{1}{8},\frac{33}{8})
   ,\qquad \ketw{\R_{2,6}^{0,2}}\sim\mathcal{E}(\frac{33}{8},\frac{1}{8})\nn
 &&\mbox{}\hspace{-1cm}\ketw{\R_{1,3}^{0,1}}\sim\mathcal{E}(1,7;0)
   ,\qquad \ketw{\R_{1,6}^{0,2}}\sim\mathcal{E}(7,1),\qquad\qquad
   \ketw{\R_{2,3}^{0,2}}\sim\mathcal{E}(\frac{5}{8},\frac{21}{8})
   ,\qquad \ketw{\R_{2,6}^{0,1}}\sim\mathcal{E}(\frac{21}{8},\frac{5}{8})\nn
 &&\mbox{}\hspace{-1cm}\ketw{\R_{1,3}^{0,2}}\sim\mathcal{E}(2,5;0)
   ,\qquad \ketw{\R_{1,6}^{0,1}}\sim\mathcal{E}(5,2)
\eea

We can encode the Jordan-cell structure of a ${\cal W}$-indecomposable rank-2
representation in its character by introducing the matrix
\be
 \mathcal{J}_2\ =\ \begin{pmatrix} 1&1 \\ 0&1 \end{pmatrix}
\ee
Its trace is simply $\mathrm{Tr}(\mathcal{J}_2)=2$  
but can be used to indicate the presence of Jordan cells of rank 2. By 
\be
 \mathrm{Tr}(\mathcal{J}_2)\big(\ch_{r,s}(q)+\ch_{r',s'}(q)\big)+2\ch_{r'',s''}(q)
\ee
we thus mean a sum of 6 irreducible characters
where a Jordan cell of rank 2 is formed between every pair of matching states
in the 2 modules labelled by $r,s$ and between every pair of matching states
in the 2 modules labelled by $r',s'$ while no state in the modules
labelled by $r'',s''$ is part of a non-trivial Jordan cell.
The characters of the ${\cal W}$-indecomposable rank-2 representations then read
\bea
 \chit[\ketw{\R_{2\kappa,s}^{1,0}}](q)&=&\delta_{\kappa,1}\big\{1-\delta_{s,3}\big\}
   +2\sum_{k\in\mathbb{N}}(2k+1-\kappa)\ch_{4k+3-2\kappa,s}(q)\nn
  &&\hspace{1.2cm} 
     +\mathrm{Tr}(\mathcal{J}_2)\sum_{k\in\mathbb{N}}(2k-2+\kappa)\ch_{4k-3+2\kappa,s}(q)  \nn
 \chit[\ketw{\R_{r,3\kappa}^{0,b}}](q)&=&\delta_{\kappa,1}\big\{1-\delta_{r,2}\big\}
   +2\sum_{k\in\mathbb{N}}(2k+1-\kappa)\ch_{r,6k+6-3\kappa-b}(q)\nn
  &&\hspace{1.2cm}
    +\mathrm{Tr}(\mathcal{J}_2)\sum_{k\in\mathbb{N}}(2k-2+\kappa)\ch_{r,6k-6+3\kappa+b}(q) 
\eea
These refined character expressions demonstrate the inequivalence of
the various representations despite the character identities (\ref{chid1}).
The relations (\ref{chrelsum}) are valid for the refined characters as well,
whereas the relations (\ref{chrel1}) are not. 
We note that the refined character expressions contain enough information to
distinguish between the different rank-2 representations. That is, the distinctions
can be made by solely emphasizing the Jordan-cell structures without further reference to
the complete embedding patterns.

Similar refinements of the rank-3 characters are possible (see below)
but not required to demonstrate inequivalence of the associated 
${\cal W}$-indecomposable rank-3 representations.
Indeed, it suffices to focus on the presence of rank-3 Jordan cells
to which end we introduce the matrix
\be
 \mathcal{J}_3\ =\ \begin{pmatrix} 1&1&0 \\ 0&1&1 \\ 0&0&1 \end{pmatrix}
\ee
with trace $\mathrm{Tr}(\mathcal{J}_3)=3$. Ignoring Jordan cells of rank 2
all together, the `semi-refined' characters of the ${\cal W}$-indecomposable rank-3 
representations then read 
\bea
 \chit[\ketw{\R_{2\kappa,3}^{1,b}}](q)&=&2+4\sum_{k\in\mathbb{N}}k\,\ch_{2k+1,b}(q)
   +4\sum_{k\in\mathbb{N}}(2k+1-\kappa)\ch_{4k+3-2\kappa,3-b}(q)\nn
  &&\hspace{1.2cm}+\big\{\mathrm{Tr}(\mathcal{J}_3)+1\big\}\sum_{k\in\mathbb{N}}(2k-2+\kappa)
     \ch_{4k-3+2\kappa,3-b}(q)
\eea
With $\kappa,b\in\mathbb{Z}_{1,2}$, these 4 semi-refined characters correspond to
4 distinct representations despite the character identities implicit in (\ref{chitRr3}).

We conclude this discussion of embedding patterns by conjecturing that the
${\cal W}$-indecomposable rank-3 representations also have embedding structures
described by the patterns in (\ref{E}). Specifically, we conjecture that
\be
 \ketw{\R_{2\kappa,3}^{1,b}}
   \sim\mathcal{E}\Big(\ketw{\R_{2\kappa,3-b}^{1,0}},\ketw{\R_{2(3-\kappa),b}^{1,0}}\Big)
   \sim\mathcal{E}\Big(\ketw{\R_{1,3\kappa}^{0,b}},\ketw{\R_{1,3(3-\kappa)}^{0,b}}\Big)
\label{RE}
\ee
where the ${\cal W}$-irreducible representations $\ketw{\D_h}$ and $\ketw{\D_v}$
have been replaced by ${\cal W}$-indecomposable rank-2 representations.
It is noted that each of the 4 rank-3 representations is thus proposed to
be viewable in two different ways. This corresponds to viewing it as
an indecomposable `vertical' combination of `horizontal' rank-2 representations
$\ketw{\R^{1,0}}$ or as an indecomposable `horizontal' combination
of `vertical' rank-2 representations $\ketw{\R^{0,b}}$.
Converting the two rank-2 Jordan cells linked by a horizontal arrow into
a rank-3 and a rank-1 Jordan cell, we finally arrive at the announced refined characters
\bea
 \chit[\ketw{\R_{2,3}^{1,1}}](q)&=&\mathrm{Tr}(\mathcal{J}_2)\chih_0(q)
   +\big\{\mathrm{Tr}(\mathcal{J}_3)+1\big\}\chih_1(q)
   +4\chih_2(q)+2\mathrm{Tr}(\mathcal{J}_2)\chih_5(q)+2\mathrm{Tr}(\mathcal{J}_2)\chih_7(q)\nn
 \chit[\ketw{\R_{4,3}^{1,1}}](q)&=&2\chih_0(q)+2\mathrm{Tr}(\mathcal{J}_2)\chih_1(q)
   +2\mathrm{Tr}(\mathcal{J}_2)\chih_2(q)+\big\{\mathrm{Tr}(\mathcal{J}_3)+1\big\}\chih_5(q)
   +4\chih_7(q)\nn
 \chit[\ketw{\R_{2,3}^{1,2}}](q)&=&\mathrm{Tr}(\mathcal{J}_2)\chih_0(q)+4\chih_1(q)
   +\big\{\mathrm{Tr}(\mathcal{J}_3)+1\big\}\chih_2(q)+2\mathrm{Tr}(\mathcal{J}_2)\chih_5(q)
   +2\mathrm{Tr}(\mathcal{J}_2)\chih_7(q)\nn
 \chit[\ketw{\R_{4,3}^{1,2}}](q)&=&2\chih_0(q)+2\mathrm{Tr}(\mathcal{J}_2)\chih_1(q)
   +2\mathrm{Tr}(\mathcal{J}_2)\chih_2(q)+4\chih_5(q)
   +\big\{\mathrm{Tr}(\mathcal{J}_3)+1\big\}\chih_7(q)
\eea
here expressed explicitly in terms of the ${\cal W}$-irreducible characters (\ref{chih}).

\subsection{Summary of ${\cal W}$-extended fusion algebra}

We denote the fusion product in the ${\cal W}$-extended picture by $\fus$ and
reserve the symbol $\otimes$ for the fusion product in the Virasoro picture.
A summary of the fusion algebra of critical percolation in the ${\cal W}$-extended picture
${\cal WLM}(2,3)$ is given in the following.
It is both associative and commutative.
To compactify the results a bit, we introduce the following linear combinations
\bea
 \Cc_s\ =\ 2\ketw{2,s}\oplus2\ketw{4,s},&\qquad&
   \Cc^{1,0}_s\ =\ 2\ketw{\R_{2,s}^{1,0}}\oplus2\ketw{\R_{4,s}^{1,0}}\nn
 \Cc^{0,b}\ =\ 2\ketw{\R_{2,3}^{0,b}}\oplus2\ketw{\R_{2,6}^{0,b}},&\qquad&
   \Cc^{1,b}\ =\ 2\ketw{\R_{2,3}^{1,b}}\oplus2\ketw{\R_{4,3}^{1,b}}\nn
 \hat\Cc^0\ =\ 4\Cc_3\oplus2\Cc^{0,1}\oplus2\Cc^{0,2},&\qquad&
   \hat\Cc^1\ =\ 4\Cc^{1,0}_3\oplus2\Cc^{1,1}\oplus2\Cc^{1,2}
\eea
and
\be
 \Dc_{\kappa,3\kappa'}^{0,b}\ =\ 2\ketw{\kappa,3(3-b\cdot\kappa')}
    \oplus2\ketw{\R_{\kappa,3\kappa'}^{0,b}},\qquad
 \Dc_{2\kappa,3}^{1,b}\ =\ 2\ketw{\R_{2(3-b\cdot\kappa),3}^{1,0}}
    \oplus2\ketw{\R_{2(b\cdot\kappa),3}^{1,b}}
\ee
where it is recalled that $\ketw{2,6}\equiv\ketw{4,3}$ and where
\be
 m\cdot n\ =\ \frac{3-(-1)^{m+n}}{2},\qquad\qquad m,n\in\mathbb{Z}
\ee
The fusion rules are listed in the tables in Figure \ref{Cayleyr1r1} through Figure \ref{Cayleyr3r3}.
They are easily combined to form a complete Cayley table as indicated in Figure \ref{SchematicCayley}.
\begin{figure}[h]
$$
\renewcommand{\arraystretch}{1.5}
\begin{array}{c||ccc}
\hat\otimes&\mathrm{rank}\ 1&\mathrm{rank}\ 2&\mathrm{rank}\ 3
\\[4pt]
\hline \hline
\rule{0pt}{14pt}
 \mathrm{rank}\ 1
   &F_{\ref{Cayleyr1r1}}&U_{\ref{Cayleyr1r23}}^T&L_{\ref{Cayleyr1r23}}^T
\\[4pt]
 \mathrm{rank}\ 2
   &U_{\ref{Cayleyr1r23}}&\big(U_{\ref{Cayleyr2r23a}}|U_{\ref{Cayleyr2r23b}}\big)
      &\big(L_{\ref{Cayleyr2r23a}}|L_{\ref{Cayleyr2r23b}}\big)^{\! T}
\\[4pt]
 \mathrm{rank}\ 3
   &L_{\ref{Cayleyr1r23}}&\big(L_{\ref{Cayleyr2r23a}}|L_{\ref{Cayleyr2r23b}}\big)&F_{\ref{Cayleyr3r3}}
\end{array}
$$
\caption{Schematic Cayley table of the ${\cal W}$-extended fusion algebra of critical percolation.
Here $F_j$ corresponds to the table given in Figure $j$ while $F_j^T$ corresponds to the transpose thereof. By $U_j$ ($L_j$) we mean the `upper' (`lower') part of the table in 
Figure $j$ while $\big(U_{\ref{Cayleyr2r23a}}|U_{\ref{Cayleyr2r23b}}\big)$ is the 
horizontal concatenation of the tables $U_{\ref{Cayleyr2r23a}}$ and $U_{\ref{Cayleyr2r23b}}$.}
\label{SchematicCayley}
\end{figure}
%
%
%
\begin{landscape}
\pagestyle{empty}
\begin{figure}
\scriptsize
$$
\renewcommand{\arraystretch}{1.5}
\begin{array}{c||cc|cc|cc|cc}
\hat\otimes&\ketw{2,1}&\ketw{4,1}&\ketw{2,2}&\ketw{4,2}&\ketw{1,3}&\ketw{1,6}&\ketw{2,3}&\ketw{4,3}
\\[4pt]
\hline \hline
\rule{0pt}{14pt}
 \ketw{2,1}&\ketw{\R_{2,1}^{1,0}}&\ketw{\R_{4,1}^{1,0}}&\ketw{\R_{2,2}^{1,0}}&\ketw{\R_{4,2}^{1,0}}
   &\ketw{2,3}&\ketw{4,3}&\ketw{\R_{2,3}^{1,0}}&\ketw{\R_{4,3}^{1,0}}
\\[4pt]
 \ketw{4,1}&\ketw{\R_{4,1}^{1,0}}&\ketw{\R_{2,1}^{1,0}}&\ketw{\R_{4,2}^{1,0}}
   &\ketw{\R_{2,2}^{1,0}}&\ketw{4,3}&\ketw{2,3}
   &\ketw{\R_{4,3}^{1,0}}&\ketw{\R_{2,3}^{1,0}}
\\[4pt]
\hline
\rule{0pt}{14pt}
 \ketw{2,2}&\ketw{\R_{2,2}^{1,0}}&\ketw{\R_{4,2}^{1,0}}
   &\ketw{\R_{2,1}^{1,0}}\oplus\ketw{\R_{2,3}^{1,0}}&\ketw{\R_{4,1}^{1,0}}\oplus\ketw{\R_{4,3}^{1,0}}
   &\ketw{\R_{2,3}^{0,1}}&\ketw{\R_{2,6}^{0,1}}&\ketw{\R_{2,3}^{1,1}}&\ketw{\R_{4,3}^{1,1}}
\\[4pt]
 \ketw{4,2}&\ketw{\R_{4,2}^{1,0}}&\ketw{\R_{2,2}^{1,0}}
   &\ketw{\R_{4,1}^{1,0}}\oplus\ketw{\R_{4,3}^{1,0}}&\ketw{\R_{2,1}^{1,0}}\oplus\ketw{\R_{2,3}^{1,0}}
   &\ketw{\R_{2,6}^{0,1}}&\ketw{\R_{2,3}^{0,1}}&\ketw{\R_{4,3}^{1,1}}&\ketw{\R_{2,3}^{1,1}}
\\[4pt]
\hline
\rule{0pt}{14pt}
 \ketw{1,3}&\ketw{2,3}&\ketw{4,3}&\ketw{\R_{2,3}^{0,1}}&\ketw{\R_{2,6}^{0,1}}
   &\ketw{1,3}\oplus\ketw{\R_{1,3}^{0,2}}&\ketw{1,6}\oplus\ketw{\R_{1,6}^{0,2}}
   &\ketw{2,3}\oplus\ketw{\R_{2,3}^{0,2}}&\ketw{4,3}\oplus\ketw{\R_{2,6}^{0,2}}
\\[4pt]
 \ketw{1,6}&\ketw{4,3}&\ketw{2,3}&\ketw{\R_{2,6}^{0,1}}&\ketw{\R_{2,3}^{0,1}}
   &\ketw{1,6}\oplus\ketw{\R_{1,6}^{0,2}}&\ketw{1,3}\oplus\ketw{\R_{1,3}^{0,2}}
   &\ketw{4,3}\oplus\ketw{\R_{2,6}^{0,2}}&\ketw{2,3}\oplus\ketw{\R_{2,3}^{0,2}}
\\[4pt]
\hline
\rule{0pt}{14pt}
 \ketw{2,3}&\ketw{\R_{2,3}^{1,0}}&\ketw{\R_{4,3}^{1,0}}&\ketw{\R_{2,3}^{1,1}}
   &\ketw{\R_{4,3}^{1,1}}&\ketw{2,3}\oplus\ketw{\R_{2,3}^{0,2}}
   &\ketw{4,3}\oplus\ketw{\R_{2,6}^{0,2}}
   &\ketw{\R_{2,3}^{1,0}}\oplus\ketw{\R_{2,3}^{1,2}}
   &\ketw{\R_{4,3}^{1,0}}\oplus\ketw{\R_{4,3}^{1,2}}
\\[4pt]
 \ketw{4,3}&\ketw{\R_{4,3}^{1,0}}&\ketw{\R_{2,3}^{1,0}}&\ketw{\R_{4,3}^{1,1}}&\ketw{\R_{2,3}^{1,1}}
   &\ketw{4,3}\oplus\ketw{\R_{2,6}^{0,2}}&\ketw{2,3}\oplus\ketw{\R_{2,3}^{0,2}}
   &\ketw{\R_{4,3}^{1,0}}\oplus\ketw{\R_{4,3}^{1,2}}&\ketw{\R_{2,3}^{1,0}}\oplus\ketw{\R_{2,3}^{1,2}}
\end{array}
$$
\caption{Cayley table of the fusions of ${\cal W}$-indecomposable rank-1 representations 
with ${\cal W}$-indecomposable rank-1 representations.}
\label{Cayleyr1r1}
\end{figure}
\end{landscape}
%
%
%
%
%
\begin{landscape}
\pagestyle{empty}
\begin{figure}
\scriptsize
$$
\renewcommand{\arraystretch}{1.5}
\begin{array}{c||cc|cc|cc|cc}
\hat\otimes&\ketw{2,1}&\ketw{4,1}&\ketw{2,2}&\ketw{4,2}&\ketw{1,3}&\ketw{1,6}&\ketw{2,3}&\ketw{4,3}
\\[4pt]
\hline \hline
\rule{0pt}{14pt}
 \ketw{\R_{2,1}^{1,0}}&\Cc_1&\Cc_1&\Cc_2&\Cc_2
   &\ketw{\R_{2,3}^{1,0}}&\ketw{\R_{4,3}^{1,0}}&\Cc_3&\Cc_3
\\[4pt]
 \ketw{\R_{4,1}^{1,0}}&\Cc_1&\Cc_1&\Cc_2&\Cc_2
   &\ketw{\R_{4,3}^{1,0}}&\ketw{\R_{2,3}^{1,0}}&\Cc_3&\Cc_3
\\[4pt]
\hline
\rule{0pt}{14pt}
 \ketw{\R_{2,2}^{1,0}}&\Cc_2&\Cc_2&\Cc_1\oplus\Cc_3&\Cc_1\oplus\Cc_3
   &\ketw{\R_{2,3}^{1,1}}&\ketw{\R_{4,3}^{1,1}}&\Cc^{0,1}&\Cc^{0,1}
\\[4pt]
 \ketw{\R_{4,2}^{1,0}}&\Cc_2&\Cc_2&\Cc_1\oplus\Cc_3&\Cc_1\oplus\Cc_3
   &\ketw{\R_{4,3}^{1,1}}&\ketw{\R_{2,3}^{1,1}}&\Cc^{0,1}&\Cc^{0,1}
\\[4pt]
\hline
\rule{0pt}{14pt}
 \ketw{\R_{2,3}^{1,0}}&\Cc_3&\Cc_3&\Cc^{0,1}&\Cc^{0,1}
   &\ketw{\R_{2,3}^{1,0}}\oplus\ketw{\R_{2,3}^{1,2}}&\ketw{\R_{4,3}^{1,0}}\oplus\ketw{\R_{4,3}^{1,2}}
   &\Cc_3\oplus\Cc^{0,2}&\Cc_3\oplus\Cc^{0,2}
\\[4pt]
 \ketw{\R_{4,3}^{1,0}}&\Cc_3&\Cc_3&\Cc^{0,1}&\Cc^{0,1}
   &\ketw{\R_{4,3}^{1,0}}\oplus\ketw{\R_{4,3}^{1,2}}&\ketw{\R_{2,3}^{1,0}}\oplus\ketw{\R_{2,3}^{1,2}}
   &\Cc_3\oplus\Cc^{0,2}&\Cc_3\oplus\Cc^{0,2}
\\[4pt]
\hline
\rule{0pt}{14pt}
 \ketw{\R_{1,3}^{0,1}}&\ketw{\R_{2,3}^{0,1}}&\ketw{\R_{2,6}^{0,1}}
   &2\ketw{2,3}\oplus\ketw{\R_{2,3}^{0,2}}&2\ketw{4,3}\oplus\ketw{\R_{2,6}^{0,2}}
   &\Dc_{1,3}^{0,1}&\Dc_{1,6}^{0,1}
   &\Dc_{2,3}^{0,1}&\Dc_{2,6}^{0,1}
\\[4pt]
 \ketw{\R_{1,6}^{0,1}}&\ketw{\R_{2,6}^{0,1}}&\ketw{\R_{2,3}^{0,1}}
   &2\ketw{4,3}\oplus\ketw{\R_{2,6}^{0,2}}&2\ketw{2,3}\oplus\ketw{\R_{2,3}^{0,2}}
   &\Dc_{1,6}^{0,1}&\Dc_{1,3}^{0,1}
   &\Dc_{2,6}^{0,1}&\Dc_{2,3}^{0,1}
\\[4pt]
\hline
\rule{0pt}{14pt}
 \ketw{\R_{1,3}^{0,2}}&\ketw{\R_{2,3}^{0,2}}&\ketw{\R_{2,6}^{0,2}}
   &2\ketw{4,3}\oplus\ketw{\R_{2,3}^{0,1}}&2\ketw{2,3}\oplus\ketw{\R_{2,6}^{0,1}}
   &\Dc_{1,6}^{0,1}&\Dc_{1,3}^{0,1}
   &\Dc_{2,6}^{0,1}&\Dc_{2,3}^{0,1}
\\[4pt]
 \ketw{\R_{1,6}^{0,2}}&\ketw{\R_{2,6}^{0,2}}&\ketw{\R_{2,3}^{0,2}}
   &2\ketw{2,3}\oplus\ketw{\R_{2,6}^{0,1}}&2\ketw{4,3}\oplus\ketw{\R_{2,3}^{0,1}}
   &\Dc_{1,3}^{0,1}&\Dc_{1,6}^{0,1}
   &\Dc_{2,3}^{0,1}&\Dc_{2,6}^{0,1}
\\[4pt]
\hline
\rule{0pt}{14pt}
 \ketw{\R_{2,3}^{0,1}}&\ketw{\R_{2,3}^{1,1}}&\ketw{\R_{4,3}^{1,1}}
   &2\ketw{\R_{2,3}^{1,0}}\oplus\ketw{\R_{2,3}^{1,2}}&2\ketw{\R_{4,3}^{1,0}}\oplus\ketw{\R_{4,3}^{1,2}}
   &\Dc_{2,3}^{0,1}&\Dc_{2,6}^{0,1}
   &\Dc_{2,3}^{1,1}&\Dc_{4,3}^{1,1}
\\[4pt]
 \ketw{\R_{2,6}^{0,1}}&\ketw{\R_{4,3}^{1,1}}&\ketw{\R_{2,3}^{1,1}}
   &2\ketw{\R_{4,3}^{1,0}}\oplus\ketw{\R_{4,3}^{1,2}}&2\ketw{\R_{2,3}^{1,0}}\oplus\ketw{\R_{2,3}^{1,2}}
   &\Dc_{2,6}^{0,1}&\Dc_{2,3}^{0,1}
   &\Dc_{4,3}^{1,1}&\Dc_{2,3}^{1,1}
\\[4pt]
\hline
\rule{0pt}{14pt}
 \ketw{\R_{2,3}^{0,2}}&\ketw{\R_{2,3}^{1,2}}&\ketw{\R_{4,3}^{1,2}}
   &2\ketw{\R_{4,3}^{1,0}}\oplus\ketw{\R_{2,3}^{1,1}}&2\ketw{\R_{2,3}^{1,0}}\oplus\ketw{\R_{4,3}^{1,1}}
   &\Dc_{2,6}^{0,1}&\Dc_{2,3}^{0,1}
   &\Dc_{4,3}^{1,1}&\Dc_{2,3}^{1,1}
\\[4pt]
 \ketw{\R_{2,6}^{0,2}}&\ketw{\R_{4,3}^{1,2}}&\ketw{\R_{2,3}^{1,2}}
   &2\ketw{\R_{2,3}^{1,0}}\oplus\ketw{\R_{4,3}^{1,1}}&2\ketw{\R_{4,3}^{1,0}}\oplus\ketw{\R_{2,3}^{1,1}}
   &\Dc_{2,3}^{0,1}&\Dc_{2,6}^{0,1}
   &\Dc_{2,3}^{1,1}&\Dc_{4,3}^{1,1}
\\[4pt]
\hline\hline
\rule{0pt}{14pt}
 \ketw{\R_{2,3}^{1,1}}&\Cc^{0,1}&\Cc^{0,1}&2\Cc_3\oplus\Cc^{0,2}&2\Cc_3\oplus\Cc^{0,2}
   &\Dc_{2,3}^{1,1}&\Dc_{4,3}^{1,1}
   &2\Cc_3\oplus2\Cc^{0,1}&2\Cc_3\oplus2\Cc^{0,1}
\\[4pt]
 \ketw{\R_{4,3}^{1,1}}&\Cc^{0,1}&\Cc^{0,1}&2\Cc_3\oplus\Cc^{0,2}&2\Cc_3\oplus\Cc^{0,2}
   &\Dc_{4,3}^{1,1}&\Dc_{2,3}^{1,1}
   &2\Cc_3\oplus2\Cc^{0,1}&2\Cc_3\oplus2\Cc^{0,1}
\\[4pt]
\hline
\rule{0pt}{14pt}
 \ketw{\R_{2,3}^{1,2}}&\Cc^{0,2}&\Cc^{0,2}&2\Cc_3\oplus\Cc^{0,1}&2\Cc_3\oplus\Cc^{0,1}
   &\Dc_{4,3}^{1,1}&\Dc_{2,3}^{1,1}
   &2\Cc_3\oplus2\Cc^{0,1}&2\Cc_3\oplus2\Cc^{0,1}
\\[4pt]
 \ketw{\R_{4,3}^{1,2}}&\Cc^{0,2}&\Cc^{0,2}&2\Cc_3\oplus\Cc^{0,1}&2\Cc_3\oplus\Cc^{0,1}
   &\Dc_{2,3}^{1,1}&\Dc_{4,3}^{1,1}
   &2\Cc_3\oplus2\Cc^{0,1}&2\Cc_3\oplus2\Cc^{0,1}
\end{array}
$$
\caption{Table of the fusions of ${\cal W}$-indecomposable rank-1 representations 
with ${\cal W}$-indecomposable rank-2 or rank-3 representations.}
\label{Cayleyr1r23}
\end{figure}
\end{landscape}
%
%
%
%
%
%
\begin{landscape}
\pagestyle{empty}
\begin{figure}
\scriptsize
$$
\renewcommand{\arraystretch}{1.5}
\begin{array}{c||cc|cc|cc}
\hat\otimes&\ketw{\R_{2,1}^{1,0}}&\ketw{\R_{4,1}^{1,0}}&\ketw{\R_{2,2}^{1,0}}
  &\ketw{\R_{4,2}^{1,0}}&\ketw{\R_{2,3}^{1,0}}&\ketw{\R_{4,3}^{1,0}}
\\[4pt]
\hline \hline
\rule{0pt}{14pt}
 \ketw{\R_{2,1}^{1,0}}&\Cc^{1,0}_1&\Cc^{1,0}_1&\Cc^{1,0}_2
   &\Cc^{1,0}_2&\Cc^{1,0}_3&\Cc^{1,0}_3
\\[4pt]
 \ketw{\R_{4,1}^{1,0}}&\Cc^{1,0}_1&\Cc^{1,0}_1&\Cc^{1,0}_2
   &\Cc^{1,0}_2&\Cc^{1,0}_3&\Cc^{1,0}_3
\\[4pt]
\hline
\rule{0pt}{14pt}
 \ketw{\R_{2,2}^{1,0}}&\Cc^{1,0}_2&\Cc^{1,0}_2&\Cc^{1,0}_1\oplus\Cc^{1,0}_3
   &\Cc^{1,0}_1\oplus\Cc^{1,0}_3&\Cc^{1,1}&\Cc^{1,1}
\\[4pt]
 \ketw{\R_{4,2}^{1,0}}&\Cc^{1,0}_2&\Cc^{1,0}_2&\Cc^{1,0}_1\oplus\Cc^{1,0}_3
   &\Cc^{1,0}_1\oplus\Cc^{1,0}_3&\Cc^{1,1}&\Cc^{1,1}
\\[4pt]
\hline
\rule{0pt}{14pt}
 \ketw{\R_{2,3}^{1,0}}&\Cc^{1,0}_3&\Cc^{1,0}_3&\Cc^{1,1}
   &\Cc^{1,1}&\Cc^{1,0}_3\oplus\Cc^{1,2}&\Cc^{1,0}_3\oplus\Cc^{1,2}
\\[4pt]
 \ketw{\R_{4,3}^{1,0}}&\Cc^{1,0}_3&\Cc^{1,0}_3&\Cc^{1,1}
   &\Cc^{1,1}&\Cc^{1,0}_3\oplus\Cc^{1,2}&\Cc^{1,0}_3\oplus\Cc^{1,2}
\\[4pt]
\hline
\rule{0pt}{14pt}
 \ketw{\R_{1,3}^{0,1}}&\ketw{\R_{2,3}^{1,1}}&\ketw{\R_{4,3}^{1,1}}
   &2\ketw{\R_{2,3}^{1,0}}\oplus\ketw{\R_{2,3}^{1,2}}
   &2\ketw{\R_{4,3}^{1,0}}\oplus\ketw{\R_{4,3}^{1,2}}&\Dc_{2,3}^{1,1}&\Dc_{4,3}^{1,1}
\\[4pt]
 \ketw{\R_{1,6}^{0,1}}&\ketw{\R_{4,3}^{1,1}}&\ketw{\R_{2,3}^{1,1}}
   &2\ketw{\R_{4,3}^{1,0}}\oplus\ketw{\R_{4,3}^{1,2}}
   &2\ketw{\R_{2,3}^{1,0}}\oplus\ketw{\R_{2,3}^{1,2}}&\Dc_{4,3}^{1,1}&\Dc_{2,3}^{1,1}
\\[4pt]
\hline
\rule{0pt}{14pt}
 \ketw{\R_{1,3}^{0,2}}&\ketw{\R_{2,3}^{1,2}}&\ketw{\R_{4,3}^{1,2}}
   &2\ketw{\R_{4,3}^{1,0}}\oplus\ketw{\R_{2,3}^{1,1}}
   &2\ketw{\R_{2,3}^{1,0}}\oplus\ketw{\R_{4,3}^{1,1}}&\Dc_{4,3}^{1,1}&\Dc_{2,3}^{1,1}
\\[4pt]
 \ketw{\R_{1,6}^{0,2}}&\ketw{\R_{4,3}^{1,2}}&\ketw{\R_{2,3}^{1,2}}
   &2\ketw{\R_{2,3}^{1,0}}\oplus\ketw{\R_{4,3}^{1,1}}
   &2\ketw{\R_{4,3}^{1,0}}\oplus\ketw{\R_{2,3}^{1,1}}&\Dc_{2,3}^{1,1}&\Dc_{4,3}^{1,1}
\\[4pt]
\hline
\rule{0pt}{14pt}
 \ketw{\R_{2,3}^{0,1}}&\Cc^{0,1}&\Cc^{0,1}&2\Cc_3\oplus\Cc^{0,2}
   &2\Cc_3\oplus\Cc^{0,2}&2\Cc_3\oplus2\Cc^{0,1}&2\Cc_3\oplus2\Cc^{0,1}
\\[4pt]
 \ketw{\R_{2,6}^{0,1}}&\Cc^{0,1}&\Cc^{0,1}&2\Cc_3\oplus\Cc^{0,2}
   &2\Cc_3\oplus\Cc^{0,2}&2\Cc_3\oplus2\Cc^{0,1}&2\Cc_3\oplus2\Cc^{0,1}
\\[4pt]
\hline
\rule{0pt}{14pt}
 \ketw{\R_{2,3}^{0,2}}&\Cc^{0,2}&\Cc^{0,2}&2\Cc_3\oplus\Cc^{0,1}
   &2\Cc_3\oplus\Cc^{0,1}&2\Cc_3\oplus2\Cc^{0,1}&2\Cc_3\oplus2\Cc^{0,1}
\\[4pt]
 \ketw{\R_{2,6}^{0,2}}&\Cc^{0,2}&\Cc^{0,2}&2\Cc_3\oplus\Cc^{0,1}
   &2\Cc_3\oplus\Cc^{0,1}&2\Cc_3\oplus2\Cc^{0,1}&2\Cc_3\oplus2\Cc^{0,1}
\\[4pt]
\hline\hline
\rule{0pt}{14pt}
 \ketw{\R_{2,3}^{1,1}}&\Cc^{1,1}&\Cc^{1,1}&2\Cc^{1,0}_3\oplus\Cc^{1,2}
   &2\Cc^{1,0}_3\oplus\Cc^{1,2}&2\Cc^{1,0}_3\oplus2\Cc^{1,1}&2\Cc^{1,0}_3\oplus2\Cc^{1,1}
\\[4pt]
 \ketw{\R_{4,3}^{1,1}}&\Cc^{1,1}&\Cc^{1,1}&2\Cc^{1,0}_3\oplus\Cc^{1,2}
   &2\Cc^{1,0}_3\oplus\Cc^{1,2}&2\Cc^{1,0}_3\oplus2\Cc^{1,1}&2\Cc^{1,0}_3\oplus2\Cc^{1,1}
\\[4pt]
\hline
\rule{0pt}{14pt}
 \ketw{\R_{2,3}^{1,2}}&\Cc^{1,2}&\Cc^{1,2}&2\Cc^{1,0}_3\oplus\Cc^{1,1}
   &2\Cc^{1,0}_3\oplus\Cc^{1,1}&2\Cc^{1,0}_3\oplus2\Cc^{1,1}&2\Cc^{1,0}_3\oplus2\Cc^{1,1}
\\[4pt]
 \ketw{\R_{4,3}^{1,2}}&\Cc^{1,2}&\Cc^{1,2}&2\Cc^{1,0}_3\oplus\Cc^{1,1}
   &2\Cc^{1,0}_3\oplus\Cc^{1,1}&2\Cc^{1,0}_3\oplus2\Cc^{1,1}&2\Cc^{1,0}_3\oplus2\Cc^{1,1}
\end{array}
$$
\caption{First part of the table of the fusions of ${\cal W}$-indecomposable rank-2 representations 
with ${\cal W}$-indecomposable rank-2 or rank-3 representations.}
\label{Cayleyr2r23a}
\end{figure}
\end{landscape}
%
%
%
%
%
%
\begin{landscape}
\pagestyle{empty}
\begin{figure}
\scriptsize
$$
\renewcommand{\arraystretch}{1.5}
\begin{array}{c||cc|cc|cc|cc}
\hat\otimes&\ketw{\R_{1,3}^{0,1}}&\ketw{\R_{1,6}^{0,1}}&\ketw{\R_{1,3}^{0,2}}&\ketw{\R_{1,6}^{0,2}}
  &\ketw{\R_{2,3}^{0,1}}&\ketw{\R_{2,6}^{0,1}}
  &\ketw{\R_{2,3}^{0,2}}&\ketw{\R_{2,6}^{0,2}}
\\[4pt]
\hline \hline
\rule{0pt}{14pt}
 \ketw{\R_{2,1}^{1,0}}&\ketw{\R_{2,3}^{1,1}}&\ketw{\R_{4,3}^{1,1}}
   &\ketw{\R_{2,3}^{1,2}}
   &\ketw{\R_{4,3}^{1,2}}&\Cc^{0,1}
   &\Cc^{0,1}&\Cc^{0,2}&\Cc^{0,2}
\\[4pt]
 \ketw{\R_{4,1}^{1,0}}&\ketw{\R_{4,3}^{1,1}}&\ketw{\R_{2,3}^{1,1}}
   &\ketw{\R_{4,3}^{1,2}}
   &\ketw{\R_{2,3}^{1,2}}&\Cc^{0,1}
   &\Cc^{0,1}&\Cc^{0,2}&\Cc^{0,2}
\\[4pt]
\hline
\rule{0pt}{14pt}
 \ketw{\R_{2,2}^{1,0}}&2\ketw{\R_{2,3}^{1,0}}\oplus\ketw{\R_{2,3}^{1,2}}
   &2\ketw{\R_{4,3}^{1,0}}\oplus\ketw{\R_{4,3}^{1,2}}
   &2\ketw{\R_{4,3}^{1,0}}\oplus\ketw{\R_{2,3}^{1,1}}
   &2\ketw{\R_{2,3}^{1,0}}\oplus\ketw{\R_{4,3}^{1,1}}&2\Cc_3\oplus\Cc^{0,2}
   &2\Cc_3\oplus\Cc^{0,2}
   &2\Cc_3\oplus\Cc^{0,1}&2\Cc_3\oplus\Cc^{0,1}
\\[4pt]
 \ketw{\R_{4,2}^{1,0}}&2\ketw{\R_{4,3}^{1,0}}\oplus\ketw{\R_{4,3}^{1,2}}
   &2\ketw{\R_{2,3}^{1,0}}\oplus\ketw{\R_{2,3}^{1,2}}
   &2\ketw{\R_{2,3}^{1,0}}\oplus\ketw{\R_{4,3}^{1,1}}
   &2\ketw{\R_{4,3}^{1,0}}\oplus\ketw{\R_{2,3}^{1,1}}&2\Cc_3\oplus\Cc^{0,2}
   &2\Cc_3\oplus\Cc^{0,2}
   &2\Cc_3\oplus\Cc^{0,1}&2\Cc_3\oplus\Cc^{0,1}
\\[4pt]
\hline
\rule{0pt}{14pt}
 \ketw{\R_{2,3}^{1,0}}&\Dc_{2,3}^{1,1}&\Dc_{4,3}^{1,1}
   &\Dc_{4,3}^{1,1}
   &\Dc_{2,3}^{1,1}&2\Cc_3\oplus2\Cc^{0,1}
   &2\Cc_3\oplus2\Cc^{0,1}
   &2\Cc_3\oplus2\Cc^{0,1}&2\Cc_3\oplus2\Cc^{0,1}
\\[4pt]
 \ketw{\R_{4,3}^{1,0}}&\Dc_{4,3}^{1,1}&\Dc_{2,3}^{1,1}&\Dc_{2,3}^{1,1}
   &\Dc_{4,3}^{1,1}&2\Cc_3\oplus2\Cc^{0,1}
   &2\Cc_3\oplus2\Cc^{0,1}
   &2\Cc_3\oplus2\Cc^{0,1}&2\Cc_3\oplus2\Cc^{0,1}
\\[4pt]
\hline
\rule{0pt}{14pt}
 \ketw{\R_{1,3}^{0,1}}&\Dc_{1,6}^{0,1}\oplus\Dc_{1,3}^{0,2}
   &\Dc_{1,3}^{0,1}\oplus\Dc_{1,6}^{0,2}
   &\Dc_{1,3}^{0,1}\oplus\Dc_{1,6}^{0,2}
   &\Dc_{1,6}^{0,1}\oplus\Dc_{1,3}^{0,2}&\Dc_{2,6}^{0,1}\oplus\Dc_{2,3}^{0,2}
   &\Dc_{2,3}^{0,1}\oplus\Dc_{2,6}^{0,2}
   &\Dc_{2,3}^{0,1}\oplus\Dc_{2,6}^{0,2}
   &\Dc_{2,6}^{0,1}\oplus\Dc_{2,3}^{0,2}
\\[4pt]
 \ketw{\R_{1,6}^{0,1}}&\Dc_{1,3}^{0,1}\oplus\Dc_{1,6}^{0,2}
   &\Dc_{1,6}^{0,1}\oplus\Dc_{1,3}^{0,2}&\Dc_{1,6}^{0,1}\oplus\Dc_{1,3}^{0,2}
   &\Dc_{1,3}^{0,1}\oplus\Dc_{1,6}^{0,2}&\Dc_{2,3}^{0,1}\oplus\Dc_{2,6}^{0,2}
   &\Dc_{2,6}^{0,1}\oplus\Dc_{2,3}^{0,2}
   &\Dc_{2,6}^{0,1}\oplus\Dc_{2,3}^{0,2}
   &\Dc_{2,3}^{0,1}\oplus\Dc_{2,6}^{0,2}
\\[4pt]
\hline
\rule{0pt}{14pt}
 \ketw{\R_{1,3}^{0,2}}&\Dc_{1,3}^{0,1}\oplus\Dc_{1,6}^{0,2}
   &\Dc_{1,6}^{0,1}\oplus\Dc_{1,3}^{0,2}&\Dc_{1,6}^{0,1}\oplus\Dc_{1,3}^{0,2}
   &\Dc_{1,3}^{0,1}\oplus\Dc_{1,6}^{0,2}&\Dc_{2,3}^{0,1}\oplus\Dc_{2,6}^{0,2}
   &\Dc_{2,6}^{0,1}\oplus\Dc_{2,3}^{0,2}
   &\Dc_{2,6}^{0,1}\oplus\Dc_{2,3}^{0,2}
   &\Dc_{2,3}^{0,1}\oplus\Dc_{2,6}^{0,2}
\\[4pt]
 \ketw{\R_{1,6}^{0,2}}&\Dc_{1,6}^{0,1}\oplus\Dc_{1,3}^{0,2}
   &\Dc_{1,3}^{0,1}\oplus\Dc_{1,6}^{0,2}&\Dc_{1,3}^{0,1}\oplus\Dc_{1,6}^{0,2}
   &\Dc_{1,6}^{0,1}\oplus\Dc_{1,3}^{0,2}&\Dc_{2,6}^{0,1}\oplus\Dc_{2,3}^{0,2}
   &\Dc_{2,3}^{0,1}\oplus\Dc_{2,6}^{0,2}
   &\Dc_{2,3}^{0,1}\oplus\Dc_{2,6}^{0,2}
   &\Dc_{2,6}^{0,1}\oplus\Dc_{2,3}^{0,2}
\\[4pt]
\hline
\rule{0pt}{14pt}
 \ketw{\R_{2,3}^{0,1}}&\Dc_{2,6}^{0,1}\oplus\Dc_{2,3}^{0,2}
   &\Dc_{2,3}^{0,1}\oplus\Dc_{2,6}^{0,2}&\Dc_{2,3}^{0,1}\oplus\Dc_{2,6}^{0,2}
   &\Dc_{2,6}^{0,1}\oplus\Dc_{2,3}^{0,2}&\Dc_{4,3}^{1,1}\oplus\Dc_{2,3}^{1,2}
   &\Dc_{2,3}^{1,1}\oplus\Dc_{4,3}^{1,2}
   &\Dc_{2,3}^{1,1}\oplus\Dc_{4,3}^{1,2}
   &\Dc_{4,3}^{1,1}\oplus\Dc_{2,3}^{1,2}
\\[4pt]
 \ketw{\R_{2,6}^{0,1}}&\Dc_{2,3}^{0,1}\oplus\Dc_{2,6}^{0,2}
   &\Dc_{2,6}^{0,1}\oplus\Dc_{2,3}^{0,2}&\Dc_{2,6}^{0,1}\oplus\Dc_{2,3}^{0,2}
   &\Dc_{2,3}^{0,1}\oplus\Dc_{2,6}^{0,2}&\Dc_{2,3}^{1,1}\oplus\Dc_{4,3}^{1,2}
   &\Dc_{4,3}^{1,1}\oplus\Dc_{2,3}^{1,2}
   &\Dc_{4,3}^{1,1}\oplus\Dc_{2,3}^{1,2}
   &\Dc_{2,3}^{1,1}\oplus\Dc_{4,3}^{1,2}
\\[4pt]
\hline
\rule{0pt}{14pt}
 \ketw{\R_{2,3}^{0,2}}&\Dc_{2,3}^{0,1}\oplus\Dc_{2,6}^{0,2}
   &\Dc_{2,6}^{0,1}\oplus\Dc_{2,3}^{0,2}&\Dc_{2,6}^{0,1}\oplus\Dc_{2,3}^{0,2}
   &\Dc_{2,3}^{0,1}\oplus\Dc_{2,6}^{0,2}&\Dc_{2,3}^{1,1}\oplus\Dc_{4,3}^{1,2}
   &\Dc_{4,3}^{1,1}\oplus\Dc_{2,3}^{1,2}
   &\Dc_{4,3}^{1,1}\oplus\Dc_{2,3}^{1,2}
   &\Dc_{2,3}^{1,1}\oplus\Dc_{4,3}^{1,2}
\\[4pt]
 \ketw{\R_{2,6}^{0,2}}&\Dc_{2,6}^{0,1}\oplus\Dc_{2,3}^{0,2}
   &\Dc_{2,3}^{0,1}\oplus\Dc_{2,6}^{0,2}&\Dc_{2,3}^{0,1}\oplus\Dc_{2,6}^{0,2}
   &\Dc_{2,6}^{0,1}\oplus\Dc_{2,3}^{0,2}&\Dc_{4,3}^{1,1}\oplus\Dc_{2,3}^{1,2}
   &\Dc_{2,3}^{1,1}\oplus\Dc_{4,3}^{1,2}
   &\Dc_{2,3}^{1,1}\oplus\Dc_{4,3}^{1,2}
   &\Dc_{4,3}^{1,1}\oplus\Dc_{2,3}^{1,2}
\\[4pt]
\hline\hline
\rule{0pt}{14pt}
 \ketw{\R_{2,3}^{1,1}}&\Dc_{4,3}^{1,1}\oplus\Dc_{2,3}^{1,2}
   &\Dc_{2,3}^{1,1}\oplus\Dc_{4,3}^{1,2}&\Dc_{2,3}^{1,1}\oplus\Dc_{4,3}^{1,2}
   &\Dc_{4,3}^{1,1}\oplus\Dc_{2,3}^{1,2}&\hat\Cc^0&\hat\Cc^0&\hat\Cc^0&\hat\Cc^0
\\[4pt]
 \ketw{\R_{4,3}^{1,1}}&\Dc_{2,3}^{1,1}\oplus\Dc_{4,3}^{1,2}
   &\Dc_{4,3}^{1,1}\oplus\Dc_{2,3}^{1,2}&\Dc_{4,3}^{1,1}\oplus\Dc_{2,3}^{1,2}
   &\Dc_{2,3}^{1,1}\oplus\Dc_{4,3}^{1,2}&\hat\Cc^0&\hat\Cc^0&\hat\Cc^0&\hat\Cc^0
\\[4pt]
\hline
\rule{0pt}{14pt}
 \ketw{\R_{2,3}^{1,2}}&\Dc_{2,3}^{1,1}\oplus\Dc_{4,3}^{1,2}
   &\Dc_{4,3}^{1,1}\oplus\Dc_{2,3}^{1,2} &\Dc_{4,3}^{1,1}\oplus\Dc_{2,3}^{1,2}
   &\Dc_{2,3}^{1,1}\oplus\Dc_{4,3}^{1,2}&\hat\Cc^0&\hat\Cc^0&\hat\Cc^0&\hat\Cc^0
\\[4pt]
 \ketw{\R_{4,3}^{1,2}}&\Dc_{4,3}^{1,1}\oplus\Dc_{2,3}^{1,2}
   &\Dc_{2,3}^{1,1}\oplus\Dc_{4,3}^{1,2}&\Dc_{2,3}^{1,1}\oplus\Dc_{4,3}^{1,2}
   &\Dc_{4,3}^{1,1}\oplus\Dc_{2,3}^{1,2}&\hat\Cc^0&\hat\Cc^0&\hat\Cc^0&\hat\Cc^0
\end{array}
$$
\caption{Second part of the table of the fusions of ${\cal W}$-indecomposable rank-2 representations 
with ${\cal W}$-indecomposable rank-2 or rank-3 representations.}
\label{Cayleyr2r23b}
\end{figure}
\end{landscape}
%
%
%
%
%
%
\begin{figure}
\scriptsize
$$
\renewcommand{\arraystretch}{1.5}
\begin{array}{c||cc|cc}
\hat\otimes&\ketw{\R_{2,3}^{1,1}}&\ketw{\R_{4,3}^{1,1}}&\ketw{\R_{2,3}^{1,2}}&\ketw{\R_{4,3}^{1,2}}
\\[4pt]
\hline \hline
\rule{0pt}{14pt}
 \ketw{\R_{2,3}^{1,1}}
   &\hat\Cc^1&\hat\Cc^1&\hat\Cc^1&\hat\Cc^1
\\[4pt]
 \ketw{\R_{4,3}^{1,1}}
   &\hat\Cc^1&\hat\Cc^1&\hat\Cc^1&\hat\Cc^1
\\[4pt]
\hline
\rule{0pt}{14pt}
 \ketw{\R_{2,3}^{1,2}}
   &\hat\Cc^1&\hat\Cc^1&\hat\Cc^1&\hat\Cc^1
\\[4pt]
 \ketw{\R_{4,3}^{1,2}}
   &\hat\Cc^1&\hat\Cc^1&\hat\Cc^1&\hat\Cc^1
\end{array}
$$
\caption{Cayley table of the fusions of ${\cal W}$-indecomposable rank-3 representations 
with ${\cal W}$-indecomposable rank-3 representations.}
\label{Cayleyr3r3}
\end{figure}
%
%
\noindent
All entries of the Cayley table of the fusions of ${\cal W}$-indecomposable rank-3 representations
provided in Figure \ref{Cayleyr3r3} are given by
\be
    \hat\Cc^1\ =\ 
     8\ketw{\R_{2,3}^{1,0}}\oplus8\ketw{\R_{4,3}^{1,0}}\oplus4\ketw{\R_{2,3}^{1,1}}
     \oplus4\ketw{\R_{4,3}^{1,1}}\oplus4\ketw{\R_{2,3}^{1,2}}
     \oplus4\ketw{\R_{4,3}^{1,2}}
\ee 
It is noted that the fusion algebra just listed does not contain an identity. We will 
discuss this further in Section \ref{SectionDiscussion}.

\section{Lattice Realization of ${\cal WLM}(2,3)$}
\label{SectionLattice}

In \cite{PRR08}, we used the infinite series of logarithmic minimal lattice models ${\cal LM}(1,p)$ 
to obtain ${\cal W}$-extended fusion rules applicable in the extended pictures ${\cal WLM}(1,p)$.
A crucial ingredient was the construction of a ${\cal W}$-invariant identity
representation $\ketw{1,1}$ defined as the infinite limit of a triple fusion of 
Virasoro-irreducible Kac representations in ${\cal LM}(1,p)$.
On the other hand, as indicated above and further discussed in Section \ref{SectionDiscussion},
there is no obvious natural candidate for an identity in the lattice realization of 
${\cal WLM}(2,3)$. It nevertheless turns out fruitful to adopt the use of 
infinite limits of triple fusions of Virasoro-irreducible Kac representations.
This also allows us to identify the various ${\cal W}$-representations with suitable limits of
Yang-Baxter integrable boundary conditions on the lattice.
Firmly based on the lattice-realization of the fundamental fusion algebra of ${\cal LM}(2,3)$, 
our fusion prescription for ${\cal WLM}(2,3)$ yields a commutative and associative fusion algebra.

\subsection{Horizontal component}

Working in the {\em fundamental} fusion algebra of critical percolation
${\cal LM}(2,3)$, as opposed to the less understood but larger {\em full\/} fusion algebra
\cite{RP0706,RP0707}, the only horizontal Kac representations at our disposal are
$\{(2k,1);\ k\in\mathbb{N}\}$. It is noted that these are all Virasoro-irreducible representations.
There are many possible triple fusions to consider of which the following one offers 
fairly straightforward access to the ${\cal W}$-extended horizontal component  
\be
 \lim_{n\to\infty}(4n,1)^{\otimes3}\ =\ \bigoplus_{k\in\mathbb{N}}2k(2k,1)
   \ =\ 2\Big(\bigoplus_{k\in\mathbb{N}}(2k-1)(4k-2,1)\Big)
   \oplus2\Big(\bigoplus_{k\in\mathbb{N}}2k(4k,1)\Big)
\label{4n}
\ee
Indeed, we now assert that this limit corresponds to the following direct sum of four 
${\cal W}$-indecomposable representations 
\be
 2\ketw{2,1}\oplus2\ketw{4,1}:=\ \lim_{n\to\infty}(4n,1)^{\otimes3}
\label{def4n}
\ee
whose decompositions in terms of Virasoro-irreducible representations read
\be
 \ketw{2,1}\ =\ \bigoplus_{k\in\mathbb{N}}(2k-1)(4k-2,1),\qquad\qquad
 \ketw{4,1}\ =\ \bigoplus_{k\in\mathbb{N}}2k(4k,1)
\label{2141}
\ee
Since the participating Virasoro representations all are of rank 1, 
the ${\cal W}$-indecomposable representations $\ketw{2,1}$ and $\ketw{4,1}$
themselves are of rank 1.

Without going into details, this separation or disentanglement of the triple
fusion into four ${\cal W}$-indecomposable representations can be made manifest
from the lattice by separating the set of link states accordingly.
Since no non-trivial Jordan cells are formed between the representations
on the right-hand side of (\ref{4n}), selecting the link states associated to either
$\ketw{2,1}$ or $\ketw{4,1}$ is a valid procedure.
When non-trivial Jordan cells are involved, on the other hand, such a selection may affect
the distribution and ranks of the cells and hence would not be valid.

Having identified $\ketw{2,1}$ and $\ketw{4,1}$, we now define the
${\cal W}$-indecomposable rank-2 representations
\be
 \ketw{\R_{2,1}^{1,0}}:=\ (2,1)\otimes\ketw{2,1},\qquad\qquad
 \ketw{\R_{4,1}^{1,0}}:=\ (2,1)\otimes\ketw{4,1}
\label{R2121}
\ee 
Their decompositions into Virasoro-indecomposable rank-2 representations are given in (\ref{Rr2}).
Of importance for the evaluation of fusion products below, we note that 
the ${\cal W}$-indecomposable representations (\ref{2141}) and (\ref{R2121}) 
have the stability properties
\be
 (4n,1)\otimes\ketw{2\kappa,1}\ =\ 2n\ketw{\R_{2(3-\kappa),1}^{1,0}},\qquad\qquad
   (4n,1)\otimes\ketw{\R_{2\kappa,1}^{1,0}}\ =\ 4n\ketw{2,1}\oplus4n\ketw{4,1}
\ee
and
\be
 (2,1)\otimes\ketw{\R_{2,1}^{1,0}}\ =\ (2,1)\otimes\ketw{\R_{4,1}^{1,0}}\ =\ 
   2\ketw{2,1}\oplus2\ketw{4,1}
\ee
As we will see in the following, there are many more such properties, but this list suffices for now.

{}From the lattice, we define the ${\cal W}$-extended fusion product $\hat\otimes$ by
\be
 \big\{2\ketw{2,1}\oplus2\ketw{4,1}\big\}\fus\ketw{A}:=\ 
  \lim_{n\to\infty}\Big(\frac{1}{2n}\Big)^3(4n,1)^{\otimes3}\otimes\ketw{A}
\ee
and obtain the fusions given in Figure \ref{hor1} where
\be
 \Ac_2\ =\ \ketw{2,1}\oplus\ketw{4,1},\qquad\qquad
 \Ac_\R\ =\ \ketw{\R_{2,1}^{1,0}}\oplus\ketw{\R_{4,1}^{1,0}}
\ee
\begin{figure}
$$
\renewcommand{\arraystretch}{1.5}
\begin{array}{c||cc|cc}
\hat\otimes&\ketw{2,1}&\ketw{4,1}&\ketw{\R_{2,1}^{1,0}}&\ketw{\R_{4,1}^{1,0}}
\\[4pt]
\hline \hline
\rule{0pt}{14pt}
 \ketw{2,1}&\ketw{\R_{2,1}^{1,0}}&\ketw{\R_{4,1}^{1,0}}&2\Ac_2&2\Ac_2
 \\[4pt]
 \ketw{4,1}&\ketw{\R_{4,1}^{1,0}}&\ketw{\R_{2,1}^{1,0}}&2\Ac_2&2\Ac_2
 \\[4pt]
\hline
\rule{0pt}{14pt}
 \ketw{\R_{2,1}^{1,0}}&2\Ac_2&2\Ac_2&2\Ac_\R&2\Ac_\R
 \\[4pt]
 \ketw{\R_{4,1}^{1,0}}&2\Ac_2&2\Ac_2&2\Ac_\R&2\Ac_\R
 \\[4pt]
\end{array}
$$
\caption{Cayley table of the purely horizontal fusion algebra.}
\label{hor1}
\end{figure}
To appreciate this, we consider the two cases $A=(2,1)$ and $A=(4,1)$ and find
\bea
  &&\hspace{-1.5cm}\big\{\ketw{2,1}\oplus\ketw{4,1}\big\}\fus\ketw{2,1}
    \ =\ \frac{1}{2}\lim_{n\to\infty}\Big(\frac{1}{2n}\Big)^3(4n,1)^{\otimes3}\otimes\ketw{2,1}\nn
  &=&\frac{1}{2}\lim_{n\to\infty}\Big(\frac{1}{2n}\Big)^2(4n,1)^{\otimes2}\otimes\ketw{\R_{4,1}^{1,0}}
  \ =\ \lim_{n\to\infty}\Big(\frac{1}{2n}\Big)(4n,1)\otimes\big\{\ketw{2,1}\oplus\ketw{4,1}\big\}\nn
  &=&\ketw{\R_{2,1}^{1,0}}\oplus\ketw{\R_{4,1}^{1,0}}
\label{214121}
\eea
and likewise
\be
   \big\{\ketw{2,1}\oplus\ketw{4,1}\big\}\fus\ketw{4,1}\ =\ \ketw{\R_{2,1}^{1,0}}\oplus\ketw{\R_{4,1}^{1,0}}
\label{214141}
\ee
We are still faced with the task of disentangling these results since the
identification of the individual fusions such as $\ketw{2,1}\fus\ketw{2,1}$ is ambiguous at this point. 
However, since
\be
 (4k-2,1)\otimes(4k'-2,1)\ =\ \bigoplus_{j=|k-k'|+1}^{k+k'-1}\R_{4j-2,1}^{1,0}
\ee
and with the Virasoro decomposition of $\ketw{2,1}$ in (\ref{2141}) in mind, it follows 
that the Virasoro decomposition of the fusion $\ketw{2,1}\fus\ketw{2,1}$ only
involves rank-2 representations of the form $\ketw{\R_{4j-2,1}^{1,0}}$.
Initially comparing this with (\ref{214121}) and subsequently with (\ref{214141}),
we conclude that
\be
 \ketw{2,1}\fus\ketw{2,1}\ =\ \ketw{4,1}\fus\ketw{4,1}\ =\ \ketw{\R_{2,1}^{1,0}},\qquad\qquad
   \ketw{2,1}\fus\ketw{4,1}\ =\ \ketw{\R_{4,1}^{1,0}}
\ee
In order to complete the Cayley table in Figure \ref{hor1}, we also need to evaluate fusions like 
\be
 \ketw{4,1}\fus\ketw{\R_{2,1}^{1,0}}\ =\ (2,1)\otimes\ketw{4,1}\fus\ketw{4,1}
   \ =\ (2,1)\otimes\ketw{\R_{2,1}^{1,0}}\ =\ 2\ketw{2,1}\oplus2\ketw{4,1}
\ee
and
\be
 \ketw{\R_{2,1}^{1,0}}\fus\ketw{\R_{4,1}^{1,0}}\ =\ (2,1)\otimes\Big(\ketw{4,1}\fus\ketw{\R_{2,1}^{1,0}}\Big)
   \ =\ 2\ketw{\R_{2,1}^{1,0}}\oplus2\ketw{\R_{4,1}^{1,0}}
\ee
The remaining fusions follow similarly.

Additional representations are obtained by fusing the ones above by the
simple vertical (Virasoro-indecomposable) Kac  representations $(1,2)$ and $(1,3)$.  
We thus define the rank-1 representations
\be
 \ketw{2\kappa,s}:=\ \ketw{2\kappa,1}\otimes(1,s)
   \ =\ \bigoplus_{k\in\mathbb{N}}(2k-2+\kappa)(2(2k-2+\kappa),s),\qquad\qquad s\in\mathbb{Z}_{2,3}
\label{2s}
\ee
and the rank-2 representations
\be
 \ketw{\R_{2\kappa,s}^{1,0}}:=\ \ketw{\R_{2\kappa,1}^{1,0}}\otimes(1,s)
   \ =\ \bigoplus_{k\in\mathbb{N}}(2k-2+\kappa)\R_{2(2k-2+\kappa),s}^{1,0}
     ,\qquad\qquad s\in\mathbb{Z}_{2,3}
\label{R2s}
\ee
Having ventured into the bulk part of the Kac table, we note the stability properties
\bea
 \ketw{2\kappa,1}\otimes(1,6n-3)\ =\ (2n-1)\ketw{2\kappa,3},&&\quad
   \ketw{2\kappa,1}\otimes(1,6n)\ =\ 2n\ketw{2(3-\kappa),3}  \nn
 \ketw{\R_{2\kappa,1}^{1,0}}\otimes(1,6n-3)\ =\ (2n-1)\ketw{\R_{2\kappa,3}^{1,0}},&&\quad 
   \ketw{\R_{2\kappa,1}^{1,0}}\otimes(1,6n)\ =\ 2n\ketw{\R_{2(3-\kappa),3}^{1,0}}
\eea

\subsection{Vertical component}

The vertical component is developed and described in much the same way as the
horizontal component above. {}From the lattice, we choose to consider 
\bea
 \!\!\!\!\!\!\!\lim_{n\to\infty}(1,6n-3)^{\otimes3}&=&\bigoplus_{k\in\mathbb{N}}(2k-1)(2k-1,1)\nn
  &=&3\Big(\bigoplus_{k\in\mathbb{N}}(2k-1)(1,6k-3)\Big)
    \oplus2\Big(\bigoplus_{k\in\mathbb{N}}2k\R_{1,6k}^{0,1}\Big)\oplus
    \Big(\bigoplus_{k\in\mathbb{N}}(2k-1)\R_{1,6k-3}^{0,2}\Big)
\label{6nm3}
\eea
Care has to be taken when disentangling this result in order to identify the ${\cal W}$-extended
representations involved. First, we observe that the conformal weights of the Virasoro representations 
in the first sum all have rational part $1/3$ while the Virasoro representations in the second
and third sums all have integer conformal weights. This allows us to separate the first sum
from the other two and we have
\be
 \ketw{1,3}\ =\ \bigoplus_{k\in\mathbb{N}}(2k-1)(1,6k-3)
\ee
Now, fusing this with $(1,3)$ gives
\be
 \ketw{1,3}\otimes(1,3)\ =\ \bigoplus_{k\in\mathbb{N}}(2k-1)\Big((1,6k-3)\oplus\R_{1,6k-3}^{0,2}\Big)
\ee
Having separated $\ketw{1,3}$ from this, we naturally identify the remaining part of the sum as 
the ${\cal W}$-extended rank-2 representation
\be
 \ketw{\R_{1,3}^{0,2}}\ =\ \bigoplus_{k\in\mathbb{N}}(2k-1)\R_{1,6k-3}^{0,2}
\ee
The second sum in (\ref{6nm3}) can now be isolated and is identified as
the ${\cal W}$-extended rank-2 representation
\be
 \ketw{\R_{1,6}^{0,1}}\ =\ \bigoplus_{k\in\mathbb{N}}2k\R_{1,6k}^{0,1}
\ee
We thus assert that the limit of the triple fusion in (\ref{6nm3}) corresponds to the
following sum of 6 ${\cal W}$-indecomposable representations
\be
 3\ketw{1,3}\oplus2\ketw{\R_{1,6}^{0,1}}\oplus\ketw{\R_{1,3}^{0,2}}:=\ \lim_{n\to\infty}(1,6n-3)^{\otimes3}
\label{def6nm3}
\ee

Here we emphasize a difference between the horizontal and vertical components.
In the horizontal case, we could perform the disentanglement in (\ref{def4n})
explicitly from the lattice by choosing the set of link states appropriately.
As already indicated in the discussion following (\ref{def4n}) and (\ref{2141}), this is
not necessarily possible when non-trivial Jordan cells are present.
One is faced with similar but more transparent complications in the Virasoro picture as well where
the indecomposable rank-2 representations $\R_{1,3k}^{0,2}$ 
cannot be constructed individually from the lattice but only in combination with
the Kac representations $(1,3k)$. To illustrate this, let us consider
\be
 (1,3)\otimes(1,3)\ =\ (1,3)\oplus\R_{1,3}^{0,2},\qquad\qquad
   \chit[\R_{1,3}^{0,2}](q)\ =\ \chit_{1,1}(q)+\chit_{1,5}(q)
\label{131313}
\ee
The Kac representations $(1,1)$, $(1,3)$ and $(1,5)$ are constructed by allowing
exactly 0, 2 or 4 defects, respectively, to propagate through the bulk of the lattice,
while the fusion $(1,3)\otimes(1,3)$ is evaluated by allowing 0, 2 or 4 defects to propagate 
through the bulk of the lattice. In the latter case, pairs of defects can be annihilated
thus yielding a block-{\em triangular} matrix realization of the transfer fusion matrix.
This block-triangularity may give rise to non-trivial Jordan cells as it does 
in the fusion $(1,3)\otimes(1,3)$. With reference to (\ref{131313}),
it is now tempting to regard the indecomposable
representation $\R_{1,3}^{0,2}$ as the result of allowing 0 or 4 defects to
propagate through the bulk. Since defects could be annihilated in quadruples, 
this would indeed give rise to a block-triangular matrix. However, it turns out that
no non-trivial Jordan cells are formed in this case implying that 
this choice of boundary condition simply corresponds to
the {\em direct} sum of the two indecomposable rank-1 representations $(1,1)$ and $(1,5)$.
As already mentioned, this phenomenon carries over to the ${\cal W}$-extended picture where
the limiting process, though, obscures the clarity of the Virasoro example just discussed.

To continue, we could apply the analysis based on (\ref{6nm3})
above to the infinite limit of the triple fusion of $(1,6n)$ 
with itself. Alternatively, we simply define the ${\cal W}$-extended rank-2 representation 
\be
 \ketw{\R_{1,3}^{0,1}}:=\ \ketw{1,3}\otimes(1,2)\ =\ \bigoplus_{k\in\mathbb{N}}(2k-1)\R_{1,6k-3}^{0,1}
\ee
and disentangle the fusions
\bea
 \ketw{\R_{1,3}^{0,2}}\otimes(1,2)&=&2\Big(\bigoplus_{k\in\mathbb{N}}2k(1,6k)\Big)
   \oplus\Big(\bigoplus_{k\in\mathbb{N}}(2k-1)\R_{1,6k-3}^{0,1}\Big)\nn
 \ketw{\R_{1,6}^{0,1}}\otimes(1,2)&=&2\Big(\bigoplus_{k\in\mathbb{N}}2k(1,6k)\Big)
   \oplus\Big(\bigoplus_{k\in\mathbb{N}}2k\R_{1,6k}^{0,2}\Big)
\eea
to identify
\be
  \ketw{1,6}\ =\ \bigoplus_{k\in\mathbb{N}}2k(1,6k)
\ee
and subsequently
\be
 \ketw{\R_{1,6}^{0,2}}\ =\ \bigoplus_{k\in\mathbb{N}}2k\R_{1,6k}^{0,2}
\ee
We thus have the stability properties
\be
 \ketw{\R_{1,3\kappa}^{0,b}}\otimes(1,2)
   \ =\ 2\ketw{1,3\kappa\cdot b}\oplus\ketw{\R_{1,3\kappa}^{0,3-b}}
\ee
with further stability properties reading
\bea
 \ketw{1,3\kappa}\otimes(1,6n-3)
   &=&(2n-1)\big\{\ketw{1,3\kappa}\oplus\ketw{\R_{1,3\kappa}^{0,2}}\big\}\nn
 \ketw{\R_{1,3\kappa}^{0,b}}\otimes(1,6n-3)
   &=&2(2n-1)\big\{\ketw{1,3(3-\kappa\cdot b)}\oplus\ketw{\R_{1,3\kappa\cdot b}^{0,1}}\big\}
\eea

In accordance with horizontal fusion, we use
\bea
 \big\{3\ketw{1,3}\oplus2\ketw{\R_{1,6}^{0,1}}\oplus\ketw{\R_{1,3}^{0,2}}\big\}\fus\ketw{A}&=&
   \lim_{n\to\infty}\Big(\frac{1}{2n}\Big)^3(1,6n-3)^{\otimes3}\otimes\ketw{A}\nn
 &=&\lim_{n\to\infty}\Big(\frac{1}{2n-1}\Big)^3(1,6n-3)^{\otimes3}\otimes\ketw{A}
\eea
when evaluating vertical fusions of ${\cal W}$-representations.
With the abbreviations
\be
 \Ac_{3\kappa}\ =\ \ketw{1,3\kappa}\oplus\ketw{\R_{1,3\kappa}^{0,2}},\qquad\qquad
   \Bc_{3\kappa}\ =\ \ketw{1,3\kappa}\oplus\ketw{\R_{1,3(3-\kappa)}^{0,1}}
\ee
and in much the same way as for the horizontal component,
this yields the fusion rules in Figure \ref{ver1a}.
%
%
\begin{figure}
\small
$$
\renewcommand{\arraystretch}{1.5}
\begin{array}{c||cc|cc|cc}
\hat\otimes&\ketw{1,3}&\ketw{1,6}&\ketw{\R_{1,6}^{0,1}}
  &\ketw{\R_{1,3}^{0,2}}&\ketw{\R_{1,3}^{0,1}}&\ketw{\R_{1,6}^{0,2}}
\\[4pt]
\hline \hline
\rule{0pt}{14pt}
 \ketw{1,3}&\Ac_3&\Ac_6&2\Bc_3&2\Bc_3&2\Bc_6&2\Bc_6
 \\[4pt]
 \ketw{1,6}&\Ac_6&\Ac_3&2\Bc_6&2\Bc_6&2\Bc_3&2\Bc_3
\\[4pt]
\hline
\rule{0pt}{14pt}
 \ketw{\R_{1,6}^{0,1}}&2\Bc_3&2\Bc_6&2\Ac_3\oplus2\Bc_3&2\Ac_3\oplus2\Bc_3
   &2\Ac_6\oplus2\Bc_6&2\Ac_6\oplus2\Bc_6
 \\[4pt]
 \ketw{\R_{1,3}^{0,2}}&2\Bc_3&2\Bc_6&2\Ac_3\oplus2\Bc_3&2\Ac_3\oplus2\Bc_3
   &2\Ac_6\oplus2\Bc_6&2\Ac_6\oplus2\Bc_6
\\[4pt]
\hline
\rule{0pt}{14pt}
 \ketw{\R_{1,3}^{0,1}}&2\Bc_6&2\Bc_3
   &2\Ac_6\oplus2\Bc_6&2\Ac_6\oplus2\Bc_6&2\Ac_3\oplus2\Bc_3&2\Ac_3\oplus2\Bc_3
\\[4pt]
 \ketw{\R_{1,6}^{0,2}}&2\Bc_6&2\Bc_3
&2\Ac_6\oplus2\Bc_6&2\Ac_6\oplus2\Bc_6&2\Ac_3\oplus2\Bc_3&2\Ac_3\oplus2\Bc_3
\end{array}
$$
\caption{Cayley table of the purely vertical fusion algebra.}
\label{ver1a}
\end{figure}
Further following the analysis of the horizontal component, we introduce the rank-1 representations 
\be
 \ketw{2,3\kappa}:=\ (2,1)\otimes\ketw{1,3\kappa}
   \ =\ \bigoplus_{k\in\mathbb{N}}(2k-2+\kappa)(2,3(2k-2+\kappa))
\ee
As required by consistency of notation, the representation $\ketw{2,3}$ defined in (\ref{2s})
must agree with this expression, and indeed it does since
\be
 \bigoplus_{k\in\mathbb{N}}(2k-1)(4k-2,3)\ =\ \bigoplus_{k\in\mathbb{N}}(2k-1)(2,6k-3)
\ee
It is likewise noted that
\be
 \ketw{2,6}\ \equiv\ \ketw{4,3}
\ee
since
\bea
 \bigoplus_{k\in\mathbb{N}}2k(2,6k)\ =\ \bigoplus_{k\in\mathbb{N}}2k(4k,3)
\eea
We also introduce the rank-2 representations
\be
 \ketw{\R_{2,3\kappa}^{0,b}}:=\ (2,1)\otimes\ketw{\R_{1,3\kappa}^{0,b}}
   \ =\ \bigoplus_{k\in\mathbb{N}}(2k-2+\kappa)\R_{2,3(2k-2+\kappa)}^{0,b}
\ee

\subsection{Combination of the two components}

Our notation implies that  
\be
 A\otimes\ketw{B}\propto \ketw{A}\otimes B
\ee
Particularly useful such relations are part of the stability properties
\bea
 \ketw{2,1}\otimes(1,3)&=&\ketw{2,3}\ =\ (2,1)\otimes\ketw{1,3}\nn
 \ketw{4,1}\otimes(1,3)&=&\ketw{4,3}\ =\ \frac{1}{2}(4,1)\otimes\ketw{1,3}\nn
 \frac{1}{2}\ketw{2,1}\otimes(1,6)&=&\ketw{2,6}\ =\ (2,1)\otimes\ketw{1,6}\nn
 \frac{1}{2}\ketw{4,1}\otimes(1,6)&=&\ketw{2,3}\ =\ \frac{1}{2}(4,1)\otimes\ketw{1,6}
\label{2113}
\eea
and
\bea
 \ketw{2,1}\otimes\R_{1,3}^{0,b}&=&\ketw{\R_{2,3}^{0,b}}
    \ =\ (2,1)\otimes\ketw{\R_{1,3}^{0,b}}\nn
 \ketw{4,1}\otimes\R_{1,3}^{0,b}&=&\ketw{\R_{2,6}^{0,b}}
    \ =\ \frac{1}{2}(4,1)\otimes\ketw{\R_{1,3}^{0,b}}\nn
 \frac{1}{2}\ketw{2,1}\otimes\R_{1,6}^{0,b}&=&\ketw{\R_{2,6}^{0,b}}
    \ =\ (2,1)\otimes\ketw{\R_{1,6}^{0,b}}\nn
 \frac{1}{2}\ketw{4,1}\otimes\R_{1,6}^{0,b}&=&\ketw{\R_{2,3}^{0,b}}
    \ =\ \frac{1}{2}(4,1)\otimes\ketw{\R_{1,6}^{0,b}}
\label{21R}
\eea
To illustrate the derivation of these, we assume (\ref{2113}) when considering the first equality 
in the third line in (\ref{21R})
\bea
 \ketw{2,1}\otimes\R_{1,6}^{0,1}&=&\ketw{2,1}\otimes(1,6)\otimes(1,2)
    \ =\ 2(2,1)\otimes\ketw{1,6}\otimes(1,2)
   \ =\ 2(2,1)\otimes\ketw{\R_{1,6}^{0,1}}\nn
 &=&2\ketw{\R_{2,6}^{0,1}}
\eea

The ${\cal W}$-indecomposable rank-3 representations can be defined by
\be
 \ketw{\R_{2\kappa,3}^{1,b}}:=\ \ketw{\R_{2\kappa,1}^{1,0}}\otimes\R_{1,3}^{0,b}
\ee
or equivalently through 
\be
 \ketw{\R_{2\kappa,3}^{1,b}}\ =\ \ketw{\R_{2,3\kappa}^{1,b}}
   \ =\ \R_{2,1}^{1,0}\otimes\ketw{\R_{1,3\kappa}^{0,b}}
\ee
They have the stability properties
\be
 \ketw{\R_{2\kappa,3}^{1,b}}\otimes(1,2)
   \ =\ 2\ketw{\R_{2\kappa\cdot b,3}^{1,0}}\oplus\ketw{\R_{2\kappa,3}^{1,3-b}}
\ee

To get started with the evaluation of combined fusions, we consider
\bea
 &&\mbox{}\hspace{-1.5cm}
   \ketw{2,1}\fus\big\{3\ketw{1,3}\oplus2\ketw{\R_{1,6}^{0,1}}\oplus\ketw{\R_{1,3}^{0,2}}\big\}
   \ =\ \lim_{n\to\infty}\Big(\frac{1}{2n-1}\Big)^3\ketw{2,1}\otimes(1,6n-3)^{\otimes3}\nn
  &=&\lim_{n\to\infty}\Big(\frac{1}{2n-1}\Big)^2\ketw{2,3}\otimes(1,6n-3)^{\otimes2}
   \ =\ \lim_{n\to\infty}\Big(\frac{1}{2n-1}\Big)^2\ketw{2,1}\otimes(1,6n-3)^{\otimes2}\otimes(1,3)\nn
  &=&\ketw{2,1}\otimes(1,3)^{\otimes3}
    \ =\ \ketw{2,1}\otimes\big\{3(1,3)\oplus\frac{1}{2}2\R_{1,6}^{0,1}\oplus\R_{1,3}^{0,2}\big\}\nn
 &=&3\ketw{2,3}\oplus2\ketw{\R_{2,6}^{0,1}}\oplus\ketw{\R_{2,3}^{0,2}}
\eea
Since the multiplicities appearing in the decomposition of the fusion
$\ketw{2,1}\fus\{3\ketw{1,3}\}$ must be divisible by 3, we find that
\be
 \ketw{2,1}\fus\ketw{1,3}\ =\ \ketw{2,3}
\ee
Using a similar argument, we then deduce that
\be
 \ketw{2,1}\fus\ketw{\R_{1,6}^{0,1}}\ =\ \ketw{\R_{2,6}^{0,1}}
\ee
and finally
\be
 \ketw{2,1}\fus\ketw{\R_{1,3}^{0,2}}\ =\ \ketw{\R_{2,3}^{0,2}}
\ee
We subsequently find
\be
 \ketw{2,1}\fus\ketw{\R_{1,3}^{0,1}}\ =\ \ketw{2,1}\fus\ketw{1,3}\otimes(1,2)\ =\ \ketw{2,3}\otimes(1,2)\ =\ 
  \ketw{\R_{2,3}^{0,1}}
\ee
and hence
\bea
 \ketw{2,1}\fus\ketw{1,6}&=&\frac{1}{2}\ketw{2,1}\fus\big\{\ketw{\R_{1,3}^{0,2}}\otimes(1,2)
   \ominus\ketw{\R_{1,3}^{0,1}}\big\}
    \ =\ \frac{1}{2}\ketw{\R_{2,3}^{0,2}}\otimes(1,2)\ominus\frac{1}{2}\ketw{\R_{2,3}^{0,1}}\nn
  &=&\frac{1}{2}(2,1)\otimes\big\{2\ketw{1,6}\oplus\ketw{\R_{1,3}^{0,1}}\big\}
    \ominus\frac{1}{2}\ketw{\R_{2,3}^{0,1}}\nn
  &=&\ketw{4,3}
\eea

The remaining fusions follow similarly or by simple applications of commutativity
and associativity. Indeed, in our final example, 
we assume that all fusions but the ones between two rank-3 representations
have been examined. Thus using commutativity, associativity and the fusion
rules appearing in Figure \ref{Cayleyr1r1} through Figure \ref{Cayleyr2r23b}, we consider
the fusion
\bea
 \ketw{\R_{2,3}^{1,1}}\fus\ketw{\R_{4,3}^{1,2}}
  &=&\ketw{\R_{2,1}^{1,0}}\fus\ketw{\R_{1,3}^{0,1}}\fus\ketw{\R_{4,3}^{1,2}}
    \ =\ \ketw{\R_{2,1}^{1,0}}\fus\left\{\Dc_{4,3}^{1,1}\oplus\Dc_{2,3}^{1,2}\right\}\nn
  &=&\ketw{\R_{2,1}^{1,0}}\fus\left\{4\ketw{\R_{2,3}^{1,0}}\oplus2\ketw{\R_{4,3}^{1,1}}
    \oplus2\ketw{\R_{4,3}^{1,2}}\right\}\nn
  &=&4\Cc_3^{1,0}\oplus2\Cc^{1,1}\oplus2\Cc^{1,2}
\eea
which is recognized as $\hat\Cc^1$, cf. Figure \ref{Cayleyr3r3}.

Self-consistency of our fusion prescription requires that the evaluation
of a given fusion product based on (\ref{def4n}) must yield the same result as the
evaluation of the same fusion product based on (\ref{6nm3}), when both methods are applicable.
This can be verified explicitly and stems from the fact that the stability properties
(\ref{2113}) and (\ref{21R}) ensure that
\bea
 &&\lim_{n\to\infty}\Big(\frac{1}{2n}\Big)^3(4n,1)^{\otimes3}
   \otimes\big\{3\ketw{1,3}\oplus2\ketw{\R_{1,6}^{0,1}}\oplus\ketw{\R_{1,3}^{0,2}}\big\}\nn
 &=&\big\{2\ketw{2,1}\oplus2\ketw{4,1}\big\}
    \fus\big\{3\ketw{1,3}\oplus2\ketw{\R_{1,6}^{0,1}}\oplus\ketw{\R_{1,3}^{0,2}}\big\}\nn
 &=& \lim_{n\to\infty}\Big(\frac{1}{2n-1}\Big)^3\big\{2\ketw{2,1}\oplus2\ketw{4,1}\big\}
   \otimes(1,6n-3)^{\otimes3}
\eea

\subsection{Fusion subalgebras}

It is noted that there are many fusion subalgebras. We have already encountered
two of them, namely the horizontal and vertical fusion algebras whose Cayley tables
are given in Figure \ref{hor1} and Figure \ref{ver1a}, respectively. 
A noteworthy six-dimensional fusion subalgebra is 
\be
 \big\langle(\Ec,1),(\Oc,1),(1,\Ec),(1,\Oc)\big\rangle
   \ =\ \big\langle(\Ec,1),(\Oc,1),(1,\Ec),(1,\Oc),(\Ec,\Ec),(\Oc,\Oc)\big\rangle
\label{EO}
\ee
It is generated by the four ${\cal W}$-representations
\be
 (\Ec,1):=\ \lim_{n\to\infty}(4n,1)^{\otimes3}\ =\ 2\Ac_2,\qquad\qquad
 (\Oc,1):=\ \frac{1}{2}(2,1)\otimes(\Ec,1)\ =\ \Ac_{\R}
\label{E1}
\ee
where it is noted that $\lim_{n\to\infty}(4n-2,1)^{\otimes3}=\lim_{n\to\infty}(4n,1)^{\otimes3}$, and
\be
 (1,\Ec):=\ \lim_{n\to\infty}(1,6n)^{\otimes3}\ =\ \Ac_6\oplus2\Bc_6,\qquad\qquad
 (1,\Oc):=\ \lim_{n\to\infty}(1,6n-3)^{\otimes3}\ =\ \Ac_3\oplus2\Bc_3
\label{1E}
\ee
The remaining two representations are defined by
\bea
 &&(\Ec,\Ec):=\ (\Ec,1)\fus(1,\Ec)\ =\ 
   \bigoplus_{\kappa\in\mathbb{Z}_{1,2},b\in\mathbb{Z}_{0,2}}\!\!\!\!(6-2b)\ketw{\R_{2,3\kappa}^{0,b}}\nn
 &&(\Oc,\Oc):=\ (\Oc,1)\fus(1,\Oc)\ =\ 
   \bigoplus_{\kappa\in\mathbb{Z}_{1,2},b\in\mathbb{Z}_{0,2}}\!\!\!\!(3-b)\ketw{\R_{2\kappa,3}^{1,b}}
\eea
where $\ketw{\R_{2,3\kappa}^{0,0}}\equiv\ketw{2,3\kappa}$,
and are seen to arise also in the fusions
\be
 (\Ec,1)\fus(1,\Oc)\ =\ (\Ec,\Ec),\qquad\qquad (\Oc,1)\fus(1,\Ec)\ =\ (\Oc,\Oc)
\ee
The Cayley table of the complete fusion subalgebra (\ref{EO}) is given in Figure \ref{CayleyEO}.
A virtue of this fusion subalgebra is that it does {\em not} rely on any disentangling procedure.
\psset{unit=1cm}
\begin{figure}
$$
\renewcommand{\arraystretch}{1.5}
\begin{array}{c||cc|cc|cc}
\hat\otimes&(\Ec,1)&(\Oc,1)&(1,\Ec)&(1,\Oc)&(\Ec,\Ec)&(\Oc,\Oc)\\[4pt]
\hline \hline
\rule{0pt}{14pt}
 (\Ec,1)&8(\Oc,1)&8(\Ec,1)&(\Ec,\Ec)&(\Ec,\Ec)&8(\Oc,\Oc)&8(\Ec,\Ec)\\[4pt]
 (\Oc,1)&8(\Ec,1)&8(\Oc,1)&(\Oc,\Oc)&(\Oc,\Oc)&8(\Ec,\Ec)&8(\Oc,\Oc)\\[4pt]
\hline
\rule{0pt}{14pt}
 (1,\Ec)&(\Ec,\Ec)&(\Oc,\Oc)&27(1,\Oc)&27(1,\Ec)&27(\Ec,\Ec)&27(\Oc,\Oc)\\[4pt]
 (1,\Oc)&(\Ec,\Ec)&(\Oc,\Oc)&27(1,\Ec)&27(1,\Oc)&27(\Ec,\Ec)&27(\Oc,\Oc)\\[4pt]
\hline
\rule{0pt}{14pt}
 (\Ec,\Ec)&8(\Oc,\Oc)&8(\Ec,\Ec)&27(\Ec,\Ec)&27(\Ec,\Ec)&216(\Oc,\Oc)&216(\Ec,\Ec)\\[4pt]
 (\Oc,\Oc)&8(\Ec,\Ec)&8(\Oc,\Oc)&27(\Oc,\Oc)&27(\Oc,\Oc)&216(\Ec,\Ec)&216(\Oc,\Oc)
\end{array}
$$
\caption{Cayley table of the $\Ec,\Oc$ fusion subalgebra.}
\label{CayleyEO}
\end{figure}

\section{Discussion}
\label{SectionDiscussion}

Two-dimensional critical percolation, with central charge $c=0$, is viewed as the member 
${\cal LM}(2,3)$ of the infinite series of Yang-Baxter integrable logarithmic minimal models 
${\cal LM}(p,p')$~\cite{PRZ}. As in the rational case~\cite{BP01}, the Yang-Baxter integrable boundary 
conditions give insight into the conformal boundary conditions~\cite{BPPZ00} 
in the continuum scaling limit as well as the fusion of their associated representations.
This enabled us in~\cite{PRZ} to construct
integrable boundary conditions labelled by $(r,s)$ and corresponding to so-called Kac representations with
conformal weights in an infinitely extended Kac table (Figure~\ref{KacTable}).
Moreover, from the lattice implementation of fusion, we obtained~\cite{RP0706} the closed fusion algebra generated
by these Kac representations finding that indecomposable representations of ranks 1, 2 and 3 are
generated by the fusion process. Although there is a countable infinity of representations,
the ensuing fusion rules are quasi-rational in the sense of Nahm~\cite{Nahm94}, that is,
the fusion of any two representations decomposes into a finite sum of representations.
This is the relevant picture in the case where the conformal algebra is the Virasoro algebra.
Of course, there is no claim, in the context of this logarithmic CFT, that the representations
generated in this picture exhaust all of the representations associated with conformal boundary conditions.
This picture is in stark contrast to the context of rational CFTs where all representations decompose into direct
sums of a finite number of irreducible representations.

In this paper, we have reconsidered critical percolation (or more precisely the ${\cal LM}(2,3)$ 
lattice model) in the continuum scaling limit to expose its nature as a `rational' logarithmic CFT
with respect to the extended conformal algebra ${\cal W}={\cal W}_{2,3}$~\cite{FGST06b}. 
Under the extended symmetry, the infinity of
Virasoro representations are reorganized into a finite number of ${\cal W}$-representations.
Following the approach of \cite{PRR08}, we construct new solutions of the boundary 
Yang-Baxter equation which, in a suitable limit, correspond to these representations. 
Specifically, with respect to a suitably defined ${\cal W}$-fusion, we find that the
representation content of the ensuing closed fusion algebra is {\em finite}
containing 26 ${\cal W}$-indecomposable representations with 8 rank-1 representations,
14 rank-2 representations and 4 rank-3 representations. We have also identified their
associated ${\cal W}$-extended characters which decompose as finite non-negative sums
of 13 ${\cal W}$-irreducible characters. Implementation of fusion on the lattice has allowed us
to read off the fusion rules governing the fusion algebra of the 26 representations and to construct an
explicit Cayley table. The closure of these representations among themselves under fusion
is remarkable confirmation of the proposed extended symmetry.

A somewhat surprising feature of our closed ${\cal W}$-extended fusion algebra of ${\cal WLM}(2,3)$ is that there appears
to be no natural identity $\Ic_{\cal W}$ expressed in terms of the fundamental Virasoro fusion
algebra and with respect to the fusion multiplication $\hat\otimes$.
Since the Kac representation $(1,1)$ is the identity
of the fundamental fusion algebra itself, it may be tempting to include it in the spectrum
and identify it with $\Ic_{\cal W}$. However, we have
\be
 \big\{2\ketw{2,1}\oplus2\ketw{4,1}\big\}\fus\Ic_{\cal W}\ =\
  \lim_{n\to\infty}\Big(\frac{1}{2n}\Big)^3(4n,1)^{\otimes3}\otimes(1,1)\ =\ 0
\ee
demonstrating that this simple extension fails.
We find it natural, though, to expect that one can extend our fusion algebra
of ${\cal WLM}(2,3)$ by working with the {\em full} Virasoro fusion algebra.
We hope to discuss this and re-address the identity question elsewhere.
\vskip.5cm
\section*{Acknowledgments}
\vskip.1cm
\noindent
This work is supported by the Australian Research Council (ARC). 
PAP acknowledges the hospitality of the Asia Pacific Center for Theoretical Physics (APCTP). 
We thank Philippe Ruelle and Ilya Tipunin for useful discussions and comments.


\end{document}